\documentclass[prx,  superscriptaddress]{revtex4-2}

\pdfoutput=1
\usepackage[caption=false]{subfig}

\usepackage{float}

\newfloat{headerbox}{!ht}

\usepackage{ragged2e}
\usepackage{wrapfig}
\usepackage{graphicx}
\usepackage{comment}
\usepackage{amsmath}
\usepackage{amssymb}
\usepackage{amsfonts}
\usepackage{bbm}
\usepackage{braket}
\usepackage{verbatim}
\usepackage[hidelinks]{hyperref}
\hypersetup{
  colorlinks   = true, 
  urlcolor     = blue, 
  linkcolor    = blue, 
  citecolor   = blue 
}
\usepackage{wrapfig}

\usepackage{color}
\usepackage{tikz}

\newcommand{\be}{\begin{equation}}
\newcommand{\ee}{\end{equation}}

\usepackage[utf8]{inputenc}
\usepackage{xcolor}
\usepackage{tikz}
\usepackage{cleveref}
\usepackage{braket}
\usepackage{titlesec}

\titleformat{\section}
  {\centering\LARGE\bfseries}{\thesection}{1em}{}

\begin{document}

\title{\huge Industry applications of neutral-atom quantum computing solving independent set problems}

\author{Jonathan Wurtz}
\affiliation{QuEra Computing Inc., 1284 Soldiers Field Road, Boston, MA, 02135, USA}

\author{Pedro Lopes}
\affiliation{QuEra Computing Inc., 1284 Soldiers Field Road, Boston, MA, 02135, USA}

\author{Christoph Gorgulla}
\affiliation{Department of Physics, Harvard University, Cambridge, MA, 02138, USA}
\affiliation{Department of Structural Biology, St Jude Children's Research Hospital, Memphis, TN, 38105, USA}

\author{Nathan Gemelke}
\affiliation{QuEra Computing Inc., 1284 Soldiers Field Road, Boston, MA, 02135, USA}

\author{Alexander Keesling}
\affiliation{QuEra Computing Inc., 1284 Soldiers Field Road, Boston, MA, 02135, USA}

\author{Sheng-Tao Wang}
\affiliation{QuEra Computing Inc., 1284 Soldiers Field Road, Boston, MA, 02135, USA}

\maketitle

Architectures for quantum computing based on neutral atoms have risen to prominence as candidates for both near and long-term applications. These devices are particularly well suited to solve independent set problems, as the combinatorial constraints can be naturally encoded in the low-energy Hilbert space due to the Rydberg blockade mechanism. Here, we approach this connection with a focus on a particular device architecture and explore the ubiquity and utility of independent set problems by providing examples of real-world applications. After a pedagogical introduction of basic graph theory concepts of relevance, we briefly discuss how to encode independent set problems in Rydberg Hamiltonians. We then outline the major classes of independent set problems and include associated example applications with industry and social relevance. We determine a wide range of sectors that could benefit from efficient solutions to independent set problems -- from telecommunications and logistics to finance and strategic planning -- and display some general strategies for efficient problem encoding and implementation on neutral-atom platforms.

\tableofcontents

\newpage

\section{Introduction}\label{sec:intro}

Quantum computers have been steadily growing in power and precision over the past few decades. Especially for the last few years, quantum computers have been more ready than ever to leap out of the academic labs: big industry players and start-ups have begun the race to turn academic proof-of-concepts into working, useful machines that may make previously impossible problems simple. Excitingly, a wide array of those problems are industry-relevant, from quantum code-breaking, encryption, and communication, to chemistry, logistics, and finance. The prospects of using today's devices to address such problems, however, are limited, due to the noisy imperfections in quantum gates (the analogue of digital logic gates) and limited numbers of qubits (the analogue of digital memory) of available machines. Yet,  with careful matchmaking, today's devices still have the potential to solve specific industry-relevant problems, and tomorrow's devices can emerge as a disruptive technology even before unlocking the full power of quantum fault-tolerance.

\begin{wrapfigure}{r}{6cm}
\includegraphics[width=6cm]{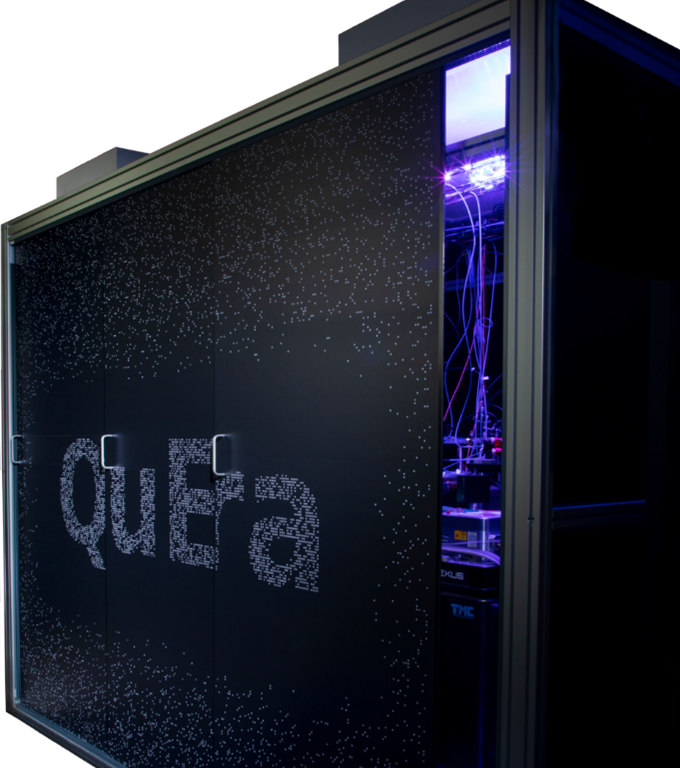}
\caption{QuEra Computing's 256-qubit neutral-atom quantum computer ``Aquila", available on Amazon Braket.}
\end{wrapfigure}

For this reason, to best exploit quantum computing today, it is imperative to understand the problems that naturally fit onto today's hardware and be able to understand the scope of the problems that can be solved with tomorrow's improvements. Unlike an abstract general-purpose quantum computer, some hardware architectures are better suited for particular kinds of problems. To this regard, QuEra is building neutral-atom quantum computers, which have yet-untapped potential to speed up the computation for a broad class of applications: the \emph{independent set problem}.

These independent set problems fall into the broad class of \emph{combinatorial optimization}, which asks: given a question with a space of possible solutions, can we find a particularly good one, as measured by the optimization of an objective function? These optimization problems are extremely ubiquitous: maximizing machine use in a factory, minimizing travel time on a delivery route, maximizing reliability in a telecommunication network, and so forth. However, these optimization problems are often computationally hard (NP-hard), which typically requires an exponential time for a classical computer to solve.

Quantum computers, however, may be able to bypass this poor performance by exploiting the quantum effects of superposition, entanglement, and interference. The neutral-atom technology developed by QuEra Computing, in particular, is well-adapted to both encode and leverage these quantum effects to solve the independent set problems~\cite{ebadi2022quantum, Pichler2018MIS}. The quantum architecture traps neutral atoms in optical tweezers and manipulates the state of the electronic orbitals (i.e.\ qubits) using focused laser beams. An atom can be kicked into a highly excited state, also called the Rydberg state, which interacts with a nearby atom, preventing them from being both excited to the Rydberg state, in a process called the \emph{Rydberg blockade}. This Rydberg blockade provides a native encoding for the independent set condition onto the array of atoms, which can be flexibly arranged in nearly arbitrary positions in one, two, or three dimensions. In conjunction with the inherent quantum coherence for the atoms, this enables efficient low-overhead programming and solution of the independent set problems via quantum algorithms, such as variational algorithms, adiabatic algorithms, quantum machine learning, and quantum sampling. Coupling quantum hardware with classical computers in a hybrid configuration, this architecture has the potential to solve the independent set problems and a large set of related problems with an efficacy that could far exceed that of classical computation. For more details on neutral-atom quantum computers and using QuEra's cloud computer ``Aquila", please refer to Ref.~\cite{wurtz2023aquila}.

\quad

Beyond the native encoding and enabling quantum solutions on QuEra's neutral-atom quantum computers, these independent set problems are both ubiquitous and useful in a wide range of industry-relevant sectors, from telecommunications and logistics, to finance and epidemiology. This paper outlines the major subcategories of independent set problems that can be addressed today with QuEra's neutral-atom quantum computers, and some example implementations of the problem to industry-specific applications. While this list is extensive, it is in no way exhaustive. Due to the ubiquity of the independent set problems, we chose a representative sample (in some ways, an ``independent set of example problems") to inspire you, dear reader, to find how your particular problems may in fact be an independent set problem that fits perfectly on QuEra's neutral-atom hardware. In this way, this paper serves to demonstrate how the independent set problems solved by QuEra's neutral-atom quantum computers can take today's hard optimization challenges and turn them into tomorrow's quantum-enabled solutions.

\quad

\quad

\quad


\begin{table}[!ht]\label{Intro_to_graphs}
    \centering
    \begin{tabular}{|c|}
    \hline
         \parbox{\textwidth}{\section{Introduction to graphs and independent sets}\label{sec:graphs-IS}
         }\\
         \hline
    \\
    \includegraphics[scale=0.35]{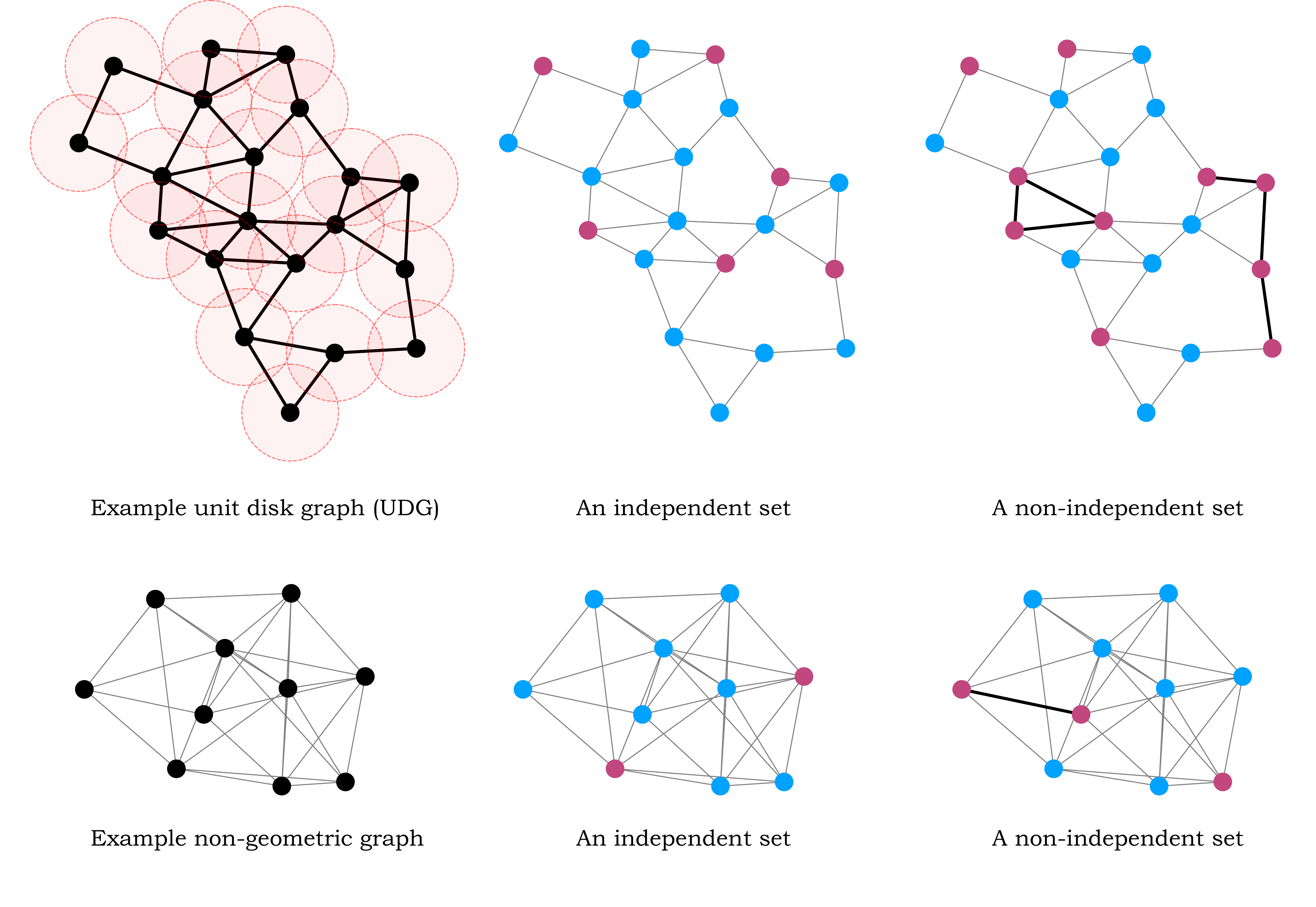}\\
    \parbox{0.9\textwidth}{\vspace{-7mm}
    \justify{\textit{Some example graphs and independent sets. A graph is a collection of vertices (blue) and edges (black) that describe relations between different objects. At the left top is a unit disk graph, where there is an edge between two vertices whenever the distance between them is less than the unit disk radii $|\vec x_i - \vec x_j|\leq R_b$. In the left bottom is a general non-geometric graph, which is simply a relation of vertices and edges. The center shows two example independent sets (red), 
    which have the constraint that no vertices in the set are connected by an edge. The right shows two non-independent sets, which have some red vertices in the set connected by some edges (dark edges), violating the independence constraint. Independent set problems broadly refer to problems that have some sort of independent set constraint.
    Many of them are hard to solve classically due to the exponential number of possible choices of independent sets, but encode many real-life problems, as will be described below.}}}\\
    \quad\\
    \hline
    \end{tabular}
\end{table}

\vspace{-8mm}
\subsection*{Combinatorial optimization
and independent set problems}

Independent set problems are constrained combinatorial optimization problems on graphs, which are generally hard for classical computers to solve. A \emph{combinatorial optimization problem} is a class of problems where one has some objective function $F(z)$ which maps some discrete domain, such as bitstrings $z \in \{0,1\}^N$, to some real number $F(z)\in \mathbb{R}$. The optimization tries to find bitstrings in this domain that maximize or minimize the objective function. This class of problems is extremely general and applies wherever discrete solutions to problems need to be optimized, which can be anything from allocating resources and optimizing portfolios, to scheduling classes and constructing robust networks. Sometimes, finding good bitstrings is easy by inspecting the structure of the function. Unfortunately, for many relevant and interesting problems, such as finding the largest independent sets, it is very ``hard" to find good bitstrings. For these problems, called \texttt{NP-hard} or non-deterministic polynomial-time hard, inspecting the structure of the problem provides no good clues and the best algorithms may resort to ``guess and check" searches of the domain by trying many solutions according to some heuristic or optimized strategy. The length of this search is often exponentially large ($2^N$) in the size of the bitstring. This may be reasonable for small-size problems; for example, for $N=32$, there are just over 4 million possible binary strings, which may be solved by a brute-force search in minutes on a laptop. However, for $N=256$, there are more possibilities ($10^{77}$)  than atoms in the universe! This exponential explosion of possible answers renders classical computers ineffective. Unfortunately, many interesting applications, including those described below, may have sizes in the hundreds or thousands, making optimization exceedingly hopeless on a classical computer.

A broad class of combinatorial optimization problems uses the language of graph theory and network analysis to define the problem. A \emph{graph} $G=(V,E)$ consists of a set of \emph{vertices} $V$, and a set of \emph{edges} $E$, and the objective function $F(z)$ can be represented in this language. The vertices may represent entities or elements of some problem, such as antennas, stocks, assets, people, locations, sensors, and so on. Each vertex may in addition have some weight $\{w_i\}$, which represents some metric of the entity, such as cost, reliability, coverage, and so on. The edges may represent relationships between different entities of the problem. For example, the vertices representing two cell phones may share an edge if they are in communication range for each other, or the vertices representing two stocks may share an edge if there is a strong correlation in their stock prices.

A subset of vertices $I\subseteq V$ is defined to be an \emph{independent set} (IS) if no two vertices in the set are adjacent to each other on the graph (i.e.~no two vertices share an edge). This is a \emph{constraint} on the discrete domain. There are $2^N$ subsets of vertices of a size-$N$ graph, while the number of independent sets is a smaller (but typically still exponential) constrained domain. The restriction has a natural interpretation for many problems, as will be discussed in the following sections. The \emph{weight} of a particular independent set is the sum of weights over each element $W_I = \sum_{i\in I} w_i$, which may represent some metric to be optimized over a particular choice of independent sets. One class of problems is called the \emph{Maximum Independent Set} (MIS) problem or the \emph{Maximum Weight Independent Set} (MWIS) problem, where the objective function $F(z) =\sum_{i\in I} w_i$ is being maximized in the constraint domain of independent sets. This optimization may correspond to, for example, the maximization of the total returns of stocks, or the minimization of the total cost of an antenna network. Here, we refer to the class of problems concerning independent sets as \emph{independent set problems}.

\vspace{-5mm}

\subsection*{Classes of graphs: general, geometric, and unit disk graphs}

\vspace{-2mm}

There are many different types of graphs. Here, we describe the features of the most relevant types.

A \emph{general graph} has no structure. It can be described simply as a set of weighted vertices $V$ connected by edges $E$ with no particular structure. To differentiate from the other broad class of graphs, we will also call these \emph{non-geometric graphs}. This could be a market graph describing the correlations of assets on a stock market, a graph of friends' interactions in a social network, and many more. An example general graph is shown in the figure above.

A \emph{geometric graph} has a spatial structure and is the complement to non-geometric graphs. For a geometric graph, each vertex has some coordinate $\{\vec x_i\}$ in some (usually 2D) space, and edges are constrained to only be between vertices that are close together. This could be a graph of high-voltage power lines between nearby cities, a graph of trade relations of different businesses in a region, and many more.

A \emph{unit disk graph} is a more constrained case of a geometric graph. As well as each vertex having some coordinate $\{\vec x_i\}$, there is an edge between every pair of vertices within some threshold distance $|\vec x_i - \vec x_j|\leq R_b$. Position can be normalized to units of $R_b$, and in 2D the distance threshold becomes a unit disk, giving the graph class its name. This could be a connection graph of in-range spatially distributed wireless nodes, zoning constraints for an urban area, and many more. An example unit disk graph is shown in the figure above.

In section \ref{sec:IS-neutral-atoms}, we describe in more detail the natural representation of independent sets of unit disk graphs on neutral-atom quantum computers with Rydberg blockade. For this reason, unit disk graphs hold a special interest, as they can be implemented on neutral-atom quantum computers with no overhead. However, we note that geometric and general graphs can still be encoded in a hybrid embedding by choosing atom locations such that their ground states map to the maximum independent set of the transformed graph (see section \ref{sec:beyondUDG}). In particular, graphs that have an underlying 2D geometric structure may still be efficiently encoded onto a 2D atom array using $\mathcal O(N)$ ancilla atoms. Any general graph may be encoded into a 2D unit disk graph using $\mathcal O(N^2)$ atoms.

Generally speaking, MIS decision problems are \texttt{NP-complete} \cite{Karp1972}, and finding maximum independent sets is \texttt{NP-hard}. Additionally, for general graphs, MIS and its reductions are \texttt{APX-complete} \cite{Hstad1999}, which means that there is no polynomial time algorithm to find good approximate solutions within a constant fraction of the exact solution. This means that in the worst case, a classical computer can do no better than picking a solution completely at random, which is generally very far from the optimal solution. Furthermore, counting the number of independent sets of a graph is \texttt{\#P-complete}~\cite{Bandyopadhyay2008, Barvinok2015}. Many other independent set problems as described in the following sections are also hard. The fact that approximate solutions are hard to find using classical computation opens up a near-term possibility for quantum speedup by using neutral-atom quantum computers to sample improved solutions.

\newpage

%
%
%
%
%
%

\begin{table}[h]\label{intro_to_computers}
\begin{tabular}[t]{|cl|}
\hline
\multicolumn{2}{|c|}{\parbox{\linewidth}{\vspace{-3mm}\section{Introduction to neutral-atom quantum computers}\vspace{-3mm}}\label{sec:neutral-atoms}} \\\hline
\parbox{0.48\linewidth}{\includegraphics[width=\linewidth]{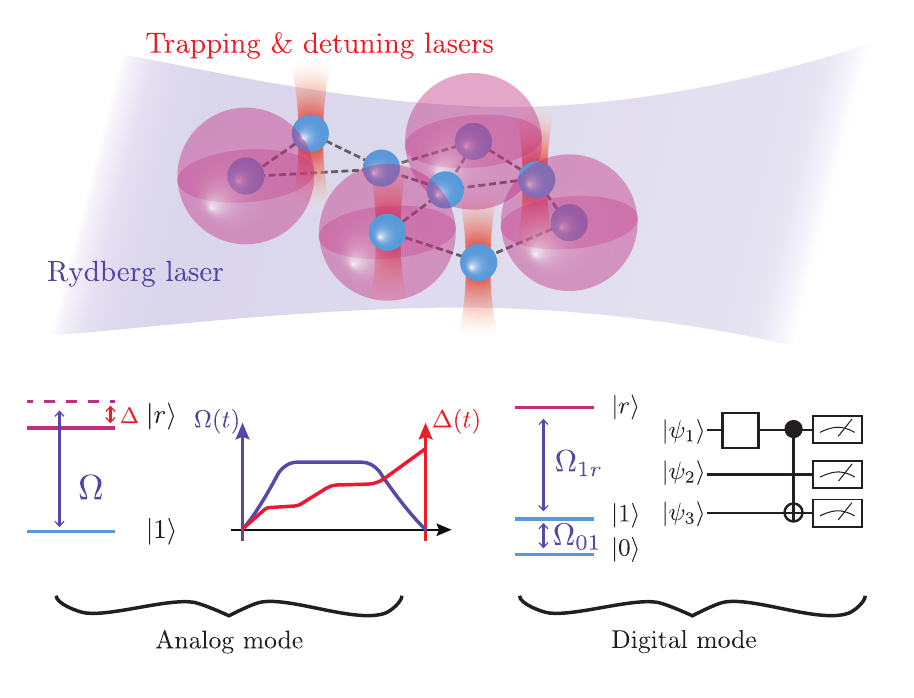}
}

&
\parbox{0.5\linewidth}{\justify{
\textbf{Top}: A representation of a neutral-atom quantum computer. Atoms (blue spheres) are trapped by optical tweezers. Each atom's internal electronic state represents the state of a qubit, as either the ground states $|0\rangle$ or $|1\rangle$, or the Rydberg state $|r\rangle$ illustrated as a large sphere. Lasers, including Rydberg lasers (purple) and local addressing lasers (red), manipulate the internal states of each atom.
\\
\textbf{Bottom Left:} Analog mode computation, which is the focus of this paper. Nearby excited atoms interact with a state-dependent van der Waals force, which shifts the energy of the Rydberg state and generates correlations and entanglement. The intensity and detuning of the lasers are manipulated to evolve the state of the atoms to encode some wavefunction, which is then measured to find solutions to hard optimization problems, such as independent set problems.
\\
\textbf{Bottom Right:} Digital mode computation, which is a future mode of QuEra's computers. The qubit is encoded in hyperfine levels $|0\rangle$ and $|1\rangle$, and lasers manipulate individual atoms to execute multi-qubit gates mediated by the Rydberg state.
}}
\\\quad&\quad\\\hline

\end{tabular}
\end{table}

Neutral-atom quantum computers have recently emerged as a promising platform for both near-term applications and long-term fault-tolerant universal quantum computation. They are particularly well suited for solving independent set problems, as the independent set constraint is naturally encoded by the interactions between Rydberg atoms. They are thus promising contenders for realizing near-term quantum speedup for solving hard combinatorial optimization problems. 

With this promise, what is a neutral-atom quantum computer? A neutral-atom quantum computer uses atoms (such as Rubidium-87) trapped by optical tweezers to construct ``nature's qubit", where the state of each qubit is encoded into the electronic orbitals of each atom. By using the natural properties of electronic levels, these qubits promise a long coherence time, enabling both near-term and potentially error-corrected quantum computation. Gates and quantum dynamics are executed using ultra-stable laser light to manipulate the electronic state of each atom. Interactions, which generate entanglement and correlations as a resource for computation, are enabled through state-dependent energy shifts of nearby atoms.

In Rydberg cold-atom devices, such as Aquila and future machines under development at QuEra, this interaction is mediated through the \emph{Rydberg state} $|r\rangle$. Such a state is a highly excited state (e.g.\ $n\sim 70$ $s$-orbital), which can be seen as a large, weakly bound ``cloud" around an atom. If two adjacent atoms are both in a Rydberg state, there is a van der Waals interaction between them, which increases the energy of the system and scales as $1/R^6$, where $R$ is the distance between the atoms. There are two modes of computation for neutral atoms: the first mode is a ``digital mode" that uses two ground states $|0\rangle$ and $|1\rangle$ to encode the qubit, which has long coherence time, and one Rydberg state $|r\rangle$ to entangle the qubits \cite{Levine_2019}. This mode can be used to execute general quantum programs and is universal. The second mode is a more limited but still powerful ``analog mode" \cite{wurtz2023aquila} that uses one ground state $|1\rangle$ and one Rydberg state $|r\rangle$, and quantum dynamics is governed by a Rydberg Hamiltonian $\hat H$ described below. Throughout this paper, we make use of the analog mode. In this mode, single qubit operations, which manipulate the atom between the ground $|1\rangle$ and Rydberg $|r\rangle$ state, are executed using detuned laser light to drive transitions to and from the Rydberg state, as shown above. Additionally, extra lasers can be used to manipulate the energy of the ground state to execute phase gates. In all, the Rydberg Hamiltonian can be written as
\begin{equation*}
\frac{\mathcal{H}}{\hbar} = \sum_j \frac{\Omega_j(t)}{2} \left( e^{i \phi_j(t) } | 1_j \rangle  \langle r_j | + e^{-i \phi_j(t) } | r_j \rangle  \langle 1_j | \right) - \sum_j \Delta_j(t) \hat n_j + \sum_{j < k} \frac{C_6}{|\overrightarrow{\mathbf{r_j}} - \overrightarrow{\mathbf{r_k}}|^6} \hat n_j \hat n_k
\end{equation*}
where $\hat n_j=|r_j\rangle\langle r_j|$ counts if the $j$th atom is in the Rydberg state, and $\Omega_j$ and $\phi_j$ are the Rabi frequency and phase of the laser driving between the ground and Rydberg state on the $j$th atom, and the last term of the Hamiltonian is the Rydberg interactions. By manipulating time-dependent fields $\Delta_j(t)$, $\Omega_j(t)$, and $\phi_j(t)$ for atom positions $\{ \overrightarrow{\mathbf{r_j}} \}$ set in arbitrary geometries, the quantum computer can prepare a state $|\psi\rangle$ as an entangled superposition of ground and Rydberg states, which can be used to solve combinatorial optimization problems. For instance, using adiabatic or variational protocols, such as the quantum approximate optimization algorithm (QAOA), one can prepare the ground state of the Hamiltonian, which may be constructed to encode the solution to independent set problems, as described in the following sections. Similarly, using variational algorithms and quantum machine learning (QML), one can construct ansatz states which provide non-classical probability distributions that may enhance classical-only optimization.

For more details on neutral-atom quantum computing, we refer to Ref.~\cite{wurtz2023aquila}, which is a comprehensive overview of the physics and theory of neutral-atom quantum computers, as well as an operating manual for implementing analog-mode quantum computations using Aquila on Amazon Braket.

\newpage

%
%
%
%
%
%

\begin{table}[!t]\label{IS_rydberg}
    \centering
    \begin{tabular}{|c|}
    \hline
\multicolumn{1}{|c|}{\parbox{\linewidth}{
\vspace{-3mm}\section{Independent set on neutral-atom quantum computers}\label{sec:IS-neutral-atoms}\vspace{-3mm}
\justify{\textbf{\LARGE{\texttt{Problem}:  Independent Set \\
\texttt{Solution}: Neutral atoms with Rydberg blockade \vspace{3mm}
}}
}}}\\
         \hline
    \quad \\
         \includegraphics[scale=0.99]{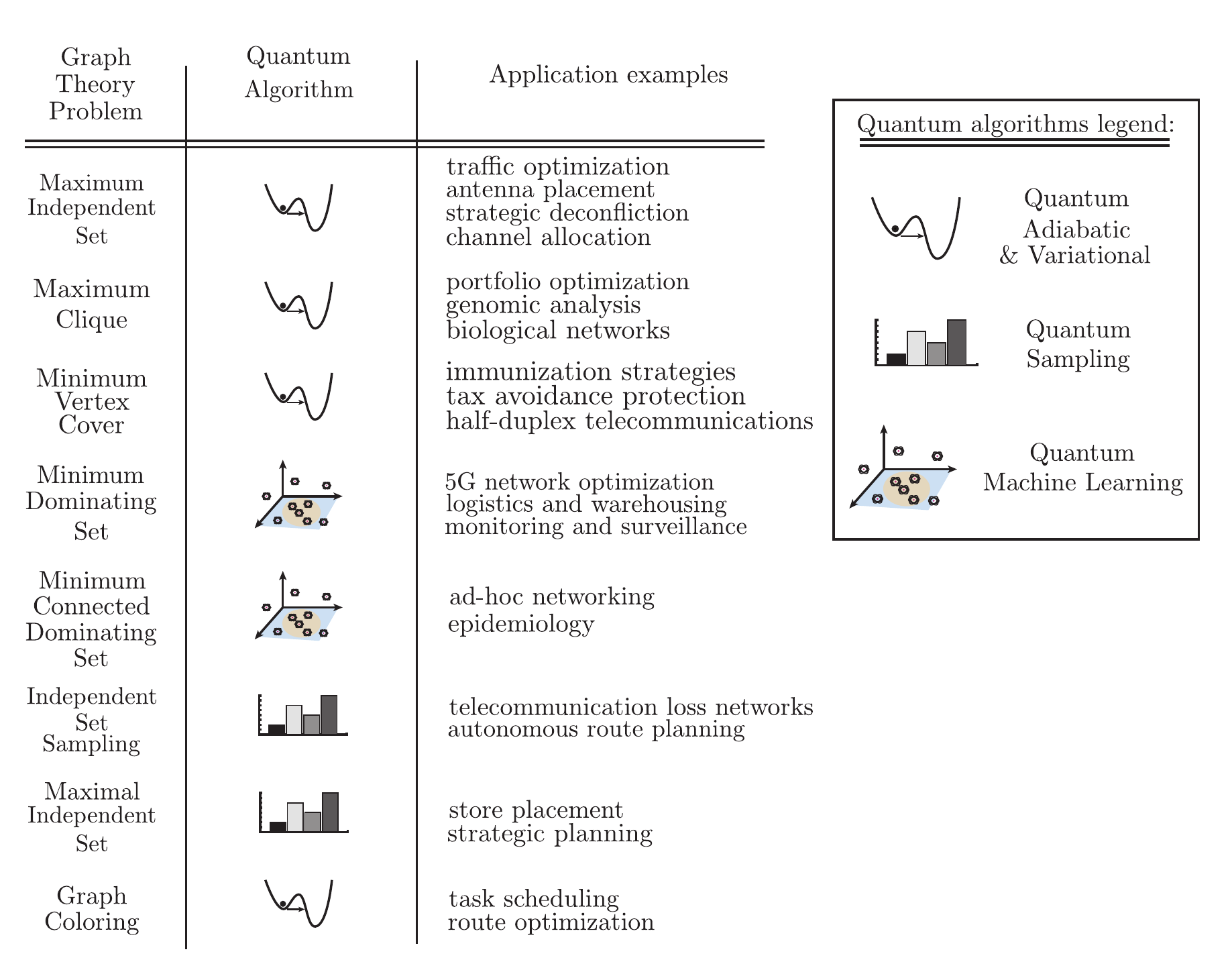}\\
    \quad \\
    \hline
    \end{tabular}
\end{table}

\newpage

While the Rydberg Hamiltonian realized by a neutral-atom quantum computer is interesting in its own right, it is also able to encode solutions to a particular class of independent set problems in its classical ground state. This is done by drawing a connection between the independence constraint and the \emph{Rydberg blockade} mechanism. For the $\Omega=0$ classical Hamiltonian with $\Delta>0$, the low energy states prefer to have as many atoms as possible excited into the Rydberg state. However, there is an energy \textit{penalty} for having two nearby atoms both in the Rydberg state, as they interact with the $1/R^6$ van der Waals force. If two atoms are within some critical radius $R_b$ where the energy of both atoms in the Rydberg state $E = -2\Delta + C_6/R^6$ is larger than the energy of a single atom in the Rydberg state $E = -\Delta$, the atoms are \emph{blockaded} from both being in the Rydberg state in a ground or low energy subspace.

It is simple to see that the Rydberg blockade is equivalent to an independence constraint on unit disk graphs, where the unit disk radius is set by the Rydberg blockade radius $R_b = (C_6/\Delta)^{1/6}$ and vertices in an independent set are labeled by atoms in a Rydberg state. The low energy subspace of the Hamiltonian then corresponds to all possible independent sets of the unit disk graph generated by the arbitrary positions of each atom in the system. In this way, the quantum state which is time evolved from one low-energy state to another (say, through an adiabatic evolution or a variational algorithm) is some entangled superposition state of independent sets, which can be an extremely powerful tool for combinatorial optimization.

There are several embeddings and algorithms that can be used to take advantage of this Rydberg blockade to solve independent set problems. The \textit{embedding} is the method by which the individual problem is encoded into the atoms, through choice of atom placement and optimization function. The \textit{algorithm} is how the wavefunction is prepared through parameterized laser pulses, as well as the classical post-processing steps that interpret measurements of the atoms' states into valid solutions.

There are two embeddings that may be used for defining an independent set problem using positions of atoms. The first is the \textit{native embedding} described in references \cite{Pichler2018MIS, ebadi2022quantum} for unit disk graphs. Here, the position of each vertex directly corresponds to the position of each atom in the array. The maximum detuning $\Delta$ and drive $\Omega$ is chosen such that the unit disk radius of the graph corresponds to the Rydberg blockade radius. Then, the low energy space is that of all independent sets of the unit disk graph. The second is the \textit{hybrid embedding}, which is possible for any graph (\textit{manuscript in preparation} and see section \ref{sec:beyondUDG}). Here, the general graph can be mapped to a unit disk graph through a gadgetization process at the cost of a polynomial (at most $N^2$) overhead of extra atoms. Then, the independent sets of the unit disk graph can be post-processed to find independent sets of the original graph. In this way, the low energy codespace of the mapped unit disk graph corresponds to all independent sets of the original, general graph.

\subsection*{Quantum adiabatic and variational algorithms}

It can be shown \cite{Pichler2018MIS} that the ground state of the classical ($\Omega=0$) Hamiltonian with $\Delta>0$ is the maximum weighted independent set of the equivalent unit disk graph. Similarly, any arbitrary graph optimization problem such as weighted independent set, clique, and others may be reduced to a maximum weighted independent set on a unit disk graph \cite{nguyen2023}. In this way, the solution of many independent set problems may be encoded in the ground state of the unit disk graph. The challenge of the quantum algorithm is then to prepare a low-energy state of the atoms, which can be measured to immediately find an answer to the optimization problem. In particular, adiabatic algorithms \cite{Albash2018} are well suited for the task. In an adiabatic algorithm, the state of a system is slowly evolved to follow the ground state of some time-dependent Hamiltonian. Beyond adiabatic protocols, variational protocols such as the quantum approximate optimization algorithm (QAOA) \cite{farhi2014quantum} or general variational ansatz \cite{Cerezo2021} may be used to generate wavefunctions that find the ground state set with an efficacy that may outperform classical computers.

\subsection*{Quantum machine learning algorithms}

Beyond simply computing the ground state of Hamiltonians which encode solutions, quantum machine learning can be used to take advantage of quantum behavior and compute optimal solutions using variational optimization and more in-depth classical post-processing. For example, if the objective function for the problem is not simply the Hamiltonian energy (such as in the dominating set examples in sections \ref{sec:dominating-set} and \ref{sec:connected-dominating-set}), the wavefunction can still be variationally optimized~\cite{farhi2017} to find good solutions. Furthermore, the probability distribution given by the quantum variational wavefunction can act as a kernel for a classical machine learning algorithm, which post-processes measurements into valid solutions. Because the probability distribution is constrained to be over independent sets of unit disk graphs, such methods may have an advantage over classical-only machine learning algorithms, as simulating such probability distributions is a \texttt{\#P}-hard counting problem.

\subsection*{Quantum sampling algorithms}

Beyond simply optimizing objective functions, the fact that the probability distribution of the wavefunction is constrained to independent sets enables powerful sampling-based machine learning algorithms. Instead of simply preparing the ground state to maximize some objective function, one may variationally prepare a state that is a superposition over many independent states, such as sampling from a classical Gibbs distribution~\cite{Wild2021}. Because the probability distribution generated by sampling the wavefunction may be biased by the variational parameters, it may generate probability distributions that are hard to reproduce classically~\cite{gao2021enhancing}. By choosing variational parameters through machine learning methods, the distribution may be fit to real-world data for analysis. Analyzing the statistics generated by these probability distributions may be useful for problems where there is uncertainty in the outcomes, as will be shown in Sec.~\ref{sec:counting-IS} and \ref{sec:maximal-IS}.

\subsection*{Outline of the paper}
Generally, being able to match the independence constraint with the Rydberg blockade constraint of neutral-atom quantum computers means that such a device is an extremely natural candidate to solve independent set problems. Starting from the long coherence properties of neutral atoms and the ability to build up entanglement and correlations simply from the structure of the underlying low-energy Hilbert space, the ability for neutral-atom quantum computers to sample from probability distributions biased to be only over independent sets enables quantum-enhanced solutions to various independent set optimization problems. As the rest of this document will describe, while this fact may be purely academic in principle, this ability can be easily applicable to a wide range of problems ranging from telecommunications to finance. Anywhere where there is a constrained optimization problem, especially for geometric data in two dimensions, QuEra's neutral-atom machines can be an excellent candidate to solve it.

\quad

The rest of this document will describe different independent set problems and detail some example applications of each. To begin, section \ref{sec:MIS} will introduce the paradigmatic optimization example of \emph{maximum independent set} (MIS). MIS is a very general problem, which may be naturally mapped to many other related problems. 
Section \ref{sec:Clique} and \ref{sec:Clique2} outlines an example of \emph{maximum clique}, which is linked to the maximum independent set through the complement graph. Similarly, section \ref{sec:vertex-cover} outlines an example of \emph{minimum vertex cover}, which is the complement vertex set to the maximum independent set. Section \ref{sec:dominating-set} outlines the \emph{minimum dominating set} problem and an application to 5G telecommunications networks. While the minimum dominating set is not naturally encoded in a ground state, it is still reasonable to use quantum machine learning algorithms and classical post-processing to find good solutions. Similarly, section \ref{sec:connected-dominating-set} extends dominating sets to connected dominating sets, a useful abstraction for ad-hoc network routing.
Section \ref{sec:counting-IS} outlines an application taking advantage of \emph{quantum sampling} from the probability distributions that are generated by quantum computers, which goes beyond the standard picture of objective function optimization. These distributions can be used, for example, for network analysis, such as finding the failure rate in a telecommunication network. In a similar theme, section \ref{sec:maximal-IS} outlines using generative modeling, classical post-processing, and optimization to sample from maximal independent sets and optimize choices under uncertainty. Section \ref{sec:graph-coloring} outlines \emph{graph coloring} and an application to the task scheduling problem. Finally, section \ref{sec:beyondUDG} outlines how unit disk graphs can be used to solve much broader problems using reductions and \emph{problem mappings}, through which the neutral-atom quantum computers can tackle a much broader class of combinatorial optimization problems.

\newpage

%
%
%
%
%
%

\begin{table}[h]\label{maximum_independent_set}
\begin{tabular}[t]{|cl|}
\hline
\multicolumn{2}{|c|}{\parbox{\linewidth}{
\vspace{-3mm}\section[Maximum independent set and an application to antenna placement]{Maximum Independent Set}\label{sec:MIS}\vspace{-3mm}
\justify{\textbf{\LARGE{
\texttt{Example:    }Antenna placement\\
\texttt{Solution: }Quantum ground state encoding}}
}}}\\ \quad & \quad\\
\hline
\parbox{0.4\linewidth}{\includegraphics[scale=2]{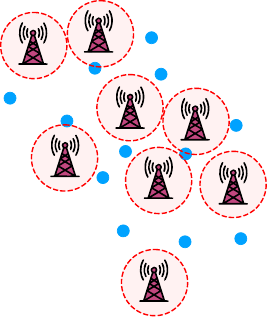}
}

&
\parbox{0.575\linewidth}{\justify{
\textit{\textbf{Left:} An example maximum independent set applied to a choice of antenna locations. When choosing locations to place antennas, it is desired to place as many antennas as possible to maximize coverage. However, antennas cannot be placed too close to each other or their signals might interfere. The optimal choice of antenna placements corresponds to a maximum independent set on a unit disk graph, where no two coverage areas overlap (red).}

\quad

\emph{Maximum independent set} (MIS) is a paradigmatic independent set problem. The MIS is the largest independent set out of all independent sets of a graph $G$. Alternatively, given a set of non-unit weights $\{w_i\}$ for each vertex, a MWIS is the independent set with the largest total weights. 
Finding maximum independent sets is a classic example of combinatorial optimization and is one of Karp's 21 NP-complete problems \cite{Karp1972}. A na\"ive algorithm relies on a brute-force search through all possible independent sets, evaluating the total weight of each one until it exhaustively finds the maximum. While special-purpose and heuristic algorithms may do better than a brute-force search, if $\texttt{P $\neq$ NP}$, then the best classical algorithms to find an MIS are (loosely) of this guess-and-check type. On the other hand, maximum independent sets are extremely useful for real-world problems. Generally, finding the maximum of a constrained optimization may be useful for a wide range of applications from profit maximization, customer coverage, network analysis, and so forth.

}}
\\\quad&\quad\\\hline

\end{tabular}
\end{table}

One example where the MIS problem is practically useful is finding antenna placement strategies that maximize coverage given distance constraints. Suppose for some region there is some set of $N$ locations $\{\vec x_i\}$ for which you may choose to place some antennae. Each antenna broadcasts and/or receives in some range $\{r_i\}$. If two antennae are too close, their signals might interfere; for this reason, it is required that two antennae are not allowed to have overlapping coverage, which occurs if they are closer than their combined range $|\vec x_i - \vec x_j|\leq r_i+r_j$. This forms a unit disk graph where vertices are antennas, and edges are between antennae with overlapping signals. This requirement of non-overlapping regions, or ``cells", forms the etymology of the phrase ``cell phone": each antenna in a region has some coverage region, which the inventors likened to biological cells. An independent set of antennae has no overlapping signals and is thus an allowed configuration. Each antenna may be given some value as a weight $w_i$, which may be, for example, the total coverage area or total number of customers that can access the network. Finding the configuration of antennae that maximizes the value of the network is thus a maximum weighted independent set (MWIS) problem.

One real-world example of this antenna placement problem is the American FM radio network, which corresponds to thousands of broadcast antenna across the United States using frequencies between 88.0 and 108.0 MHz and is regulated by the federal communications commission (FCC). Each station has a particular power to broadcast to the surrounding area, and FCC regulations require that stations on the same frequency have a minimum separation \cite{fcc_station_placement}. Additionally, the FCC requires a minimum separation distance between stations broadcasting on adjacent channels, to reduce overlap. In this way, allowed configurations of FM radio stations correspond to an independent set in a quasi-2D geometry: The 2D plane indexes the spatial locations of each station, which is duplicated across the 100 FM channels. Independence constraints are between stations on the same channel, and stations on adjacent channels, which are within the FCC-regulated overlap distance. An independent set is then the set of channels that each station is allowed to broadcast on, and a maximum independent set is the maximum utilization of the network within a region, which maximizes coverage or customers served.

\quad

There are many other applications of MIS. In fact, most of the examples shown in the rest of the document also have some use for maximum independent sets. Due to the ubiquity of MIS \cite{butenko2003maximum}, we list some other applications here, which are of course not meant to be exhaustive.

The route interaction graph of page \pageref{loss_networks} may be generalized for many problems in collision avoidance and route scheduling. For example, routes of some autonomous vehicles may overlap, risking a collision if they occupy the same area at the same time. Thus, the maximum independent set of the route interaction graph gives the largest number of routes that can simultaneously be occupied without risk of collision. As the applications of autonomous vehicles grow, this MIS analysis may be very beneficial to maximize the use of airspace in congested areas.

While the store placement problem of page \pageref{store_placement} was introduced in the context of a \texttt{\#P}-counting problem, it is extremely natural to view it as an MIS problem, where one finds a configuration that maximizes the number of locations or stores in an area. Beyond buildings, MIS on geometric graphs is extremely ubiquitous, from placing a maximum number of labels on a map \cite{agarwal1998label} to optimizing windmill placement to minimize wind shadowing effects. MIS may be useful for problems in warehousing as well: by choosing which warehouses a product is stocked in, one may maximize coverage and reliability in an area while minimizing redundancy and inventory.

While the portfolio optimization problem of page \pageref{portfolio_optimization} was introduced in the context of maximum clique, such challenges also have a natural problem statement in terms of MIS. By avoiding strongly correlated stocks or assets, one may find the MWIS to maximize return and minimize risk~\cite{florescu2016handbook}. Beyond stocks or assets, choosing strongly anticorrelated items may be useful in other contexts. As one simple example, choosing a codespace from a choice of pseudorandom signals may be useful for constructing communication channels that are robust under noise (see, for example,  \href{https://en.wikipedia.org/wiki/Code-division_multiple_access}{CDMA}) for telecommunications. As a completely different example, identifying suppliers along the supply chain who use different source materials may reduce supply chain disruptions and bottlenecks.

While the network immunization problem of page \pageref{network_immunization} was introduced in the context of vertex cover, analyzing maximum independent sets may be useful for network analysis of large interconnected systems. For example, analyzing the dependencies of different nodes of the electrical grid might inform how a failure in one part of the system may spread to different parts of the system. By hardening certain nodes, this may increase the robustness of the system at low expense. In some sense, this is still a network immunization problem, but using MIS to inform network analysis optimization may be useful to reduce failure modes. Similarly, the MIS can be used as a measure of graph isomorphisms, which is useful for applications such as molecular similarity \cite{Hernandez2016}.

Finally, a very interesting problem covered on page \pageref{task_scheduling} is scheduling, where assets must be assigned to tasks without being double-booked. In some sense, this is the same problem as collision avoidance, but in a broader context. Choosing a maximum independent set of tasks to be done optimizes the utilization of assets and increases efficiency in different contexts. For example, these tasks could be production lines on a factory floor; the choice of employees going to different meetings at a conference; choosing which trucks are used to move which items in a logistic network; and many more.

As can be seen, MIS may be applied in many ways, but there are three major themes above. The first is \emph{finding anticorrelated elements} in a network, such as stocks in a portfolio or suppliers in a supply chain. The second theme is \emph{collision avoidance}, where certain solutions are disallowed due to overlaps, such as drones occupying the same area or two stores being too close together. The final theme is \emph{network performance}, where the MIS or large independent sets inform how the network behaves under stress, such as analyzing power grid dependencies or disease spread pathways. Generally, wherever one or more of these themes comes up in a problem, it may be linked to the maximum independent set or related problems.

\subsection*{Quantum implementations}

There are many ways to compute maximum independent sets of a graph using neutral-atom quantum computers. The most natural implementation is adiabatic quantum computing \cite{Albash2018}, where the state of a system is slowly evolved to follow the ground state of some time-dependent Hamiltonian. The ground state of the Hamiltonian may be chosen to encode the maximum independent set of a graph in an extremely natural manner using the Rydberg blockade and detuning fields, and so measuring the final wavefunction of the time-evolved system encodes the maximum independent set of the problem. Beyond adiabatic protocols, variational protocols such as the QAOA~\cite{farhi2014quantum}, general variational ansatz~\cite{Cerezo2021}, and quantum machine learning (QML) \cite{Biamonte2017} may be used to generate wavefunctions which encode the maximum independent set with an efficacy that may outperform classical computers. Indeed, a demonstration at Harvard in collaboration with QuEra shows that neutral-atom quantum processors can solve the MIS problem with a superlinear quantum speedup compared to a class of generic classical algorithms~\cite{ebadi2022quantum}.

\newpage

%
%
%
%
%
%
\begin{table}[!ht]\label{portfolio_optimization}
    \centering
    \begin{tabular}{|c|}
    \hline
         \parbox{\linewidth}{
         \vspace{-3mm}\section[Maximum clique and an application to portfolio optimization]{Maximum Clique}\label{sec:Clique}\vspace{-3mm}
         \justify{\textbf{\LARGE{
\texttt{Example}:\quad Portfolio optimization\\
\texttt{Solution}:\quad Quantum ground state encoding}}
}}

\\\quad \\
         \hline
    \includegraphics[scale=2]{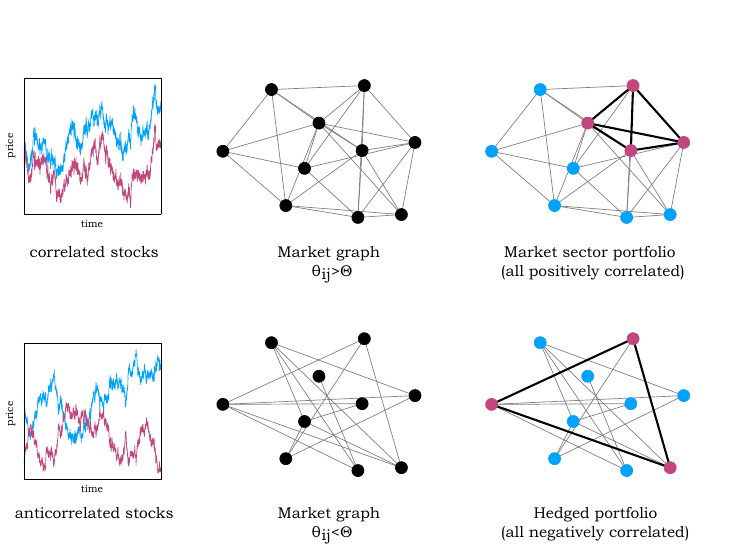}\\
    \parbox{0.9\linewidth}{ \justify{
    \textit{\textbf{Top:} An example portfolio optimization problem. Each stock has some time-dependent returns, which may be correlated (top) or anticorrelated (bottom) with other stocks. Each vertex in a market graph (middle) is a stock, where there is an edge between vertices if the assets are correlated or anticorrelated within some threshold. Cliques are market sector portfolios (top) where all stocks are correlated, or hedged portfolios (bottom) where all stocks are anticorrelated and risk is minimized. Maximum weighted cliques maximize returns under the clique constraint, which reduces variance or finds the largest market sectors.
    }
    
    \quad
    
    A \emph{maximum clique} of a graph is the largest subset of vertices that are fully connected \cite{bomze1999maximum}. This means a clique forms a complete subgraph, and every vertex is adjacent to every other vertex in the clique. The maximum clique is the natural complement to MIS. A complement graph $\overline G$ is the graph that includes every vertex in $G$, and every edge between vertices that is \emph{not} in $G$. Because no two vertices in an independent set are adjacent, every vertex in the independent set is connected in the complement graph, forming a clique. Thus, the MIS vertex set $I$ of a graph $G$ is equivalent to the maximum clique of the complement graph $\overline G$, and vice versa. \\
    
    }} \\
    \hline
    \end{tabular}
\end{table}

Portfolio optimization is one of the interesting applications of the maximum clique problem~\cite{BOGINSKI2005, florescu2016handbook}. Here, the question is in choosing a discrete number of stocks or assets that maximizes profit or returns, while minimizing the risk (e.g.\ variance of returns).
As a preprocessing step, bulk data describing the time-dependent returns of each stock is analyzed for cross-correlations within some time period. This is useful because if two stocks are anticorrelated, the combined portfolio has less variance, hence minimizing risk.

With this correlation matrix $\theta_{ij}$, one can then construct a \emph{market graph}, where vertices are individual assets weighted by return, and there is an edge between vertices if the assets' correlation exceeds some threshold. For a correlation $\theta_{ij}$ and threshold $\Theta$, the graph of edges $\theta_{ij}\leq \Theta\leq 0$ indicates stocks are anticorrelated if they are connected. Alternatively, the complement graph of edges $\theta_{ij}\geq \Theta\geq 0$ indicates stocks are correlated if they are connected. Finally, the graph of edges $|\theta_{ij}|<\Theta$ indicates that stocks are uncorrelated if they are connected. In this way, various market structures and portfolios can be found by varying the threshold $\Theta$ for the market and its complement.

Cliques of the market graph form useful portfolios. For example, for a graph with threshold $\theta_{ij}\leq \Theta\leq 0$, assets are adjacent if they are anticorrelated beyond the threshold. A clique then corresponds to a collection of assets that are all mutually anticorrelated to each other, which means that the associated portfolio will have a low risk. The maximum weighted clique then corresponds to a collection of assets that maximizes return and minimizes variance. In this way, choosing a maximum clique of the negatively correlated market graph is a \emph{hedging}, i.e., choosing assets that have complementary behavior to minimize risk.

Different-sized optimal hedging portfolios may be chosen by varying the threshold $\Theta$. For $\Theta$ large and negative, the graph connectivity will be small and so only stocks with a large anticorrelation form a clique. For this reason, the corresponding portfolio will be small. Conversely, if $\Theta$ is small, or positive, the graph connectivity will be large so the largest clique will also be larger, with a correspondingly large portfolio.

For a graph with a threshold $|\theta_{ij}|<\Theta$, each connected vertex is explicitly \textit{un}correlated within the threshold. Cliques of stocks thus behave completely independently from each other, and the maximum clique is the largest such portfolio of stocks. In this way, choosing a maximum clique of the uncorrelated market graph is a \emph{diversification} of assets, which strives to minimize risk by investing in many uncorrelated market sectors.

For a graph with a threshold $\theta_{ij}\geq\Theta\geq 0$, assets are adjacent if they are correlated beyond the threshold. A clique then corresponds to a collection of assets that are mutually correlated to each other, which means that the price of each asset follows the others. Usually, this is the case if all of the assets are part of the same market sector (say, airlines or banking). The maximum weighted clique then corresponds to a collection of assets that maximizes returns, while subject to the requirement that each asset be correlated within a threshold. In this way, choosing a maximum clique of the positively correlated market graph is a portfolio of assets within a market sector, where instead of choosing a single company, one can choose a hedged portfolio of companies within the same sector.

In the same way as the anticorrelated market graph, different-sized sector portfolios may be chosen by varying the threshold $\Theta$. Additionally, it may be useful to find large maximal cliques, which may identify different sectors of the market and thus provide an ensemble of close-to-optimal diversified portfolios across different market sectors. Furthermore, the graph coloring problem as described in section \ref{sec:graph-coloring} may be useful for partitioning the market into different sectors. Given a threshold $\Theta$ and $K$ colors corresponding to $K$ market sectors, a $K$-coloring of the positively correlated market graph corresponds to an identification of each asset with each market sector.

\quad

In general, cliques and independent sets may be useful to generate nontrivial insight in large datasets with thousands of discrete elements. For example, analyzing correlations of genome mapping data may inform the structure of biological networks~\cite{BUTENKO20061} or infer information about protein structure graphs \cite{depolli2013exact}. If the data can be pre-processed classically to generate a graph-like object, maximum clique may be used to analyze the structure of the graph and find interesting relationships between elements and thus inform scientific or business decisions.

\subsection*{Quantum implementations}
There are many ways to compute the maximum clique of a graph using neutral-atom quantum computers, by finding the maximum independent set of the complement graph. The most natural implementation is adiabatic quantum computing \cite{Albash2018}, where the state of a system is slowly evolved to follow the ground state of some time-dependent Hamiltonian. The ground state of the Hamiltonian may be chosen to encode the maximum independent set of a graph in an extremely natural manner using the Rydberg blockade and detuning fields, and so measuring the final wavefunction of the time-evolved system encodes the maximum independent set of the problem. If the problem cannot be encoded into a unit disk graph as may be the case for non-geometric graphs, it can alternatively be reduced to a unit disk graph using extra ancillary qubits~\cite{nguyen2023}. Beyond adiabatic protocols, variational protocols such as the QAOA \cite{farhi2014quantum}, general variational ansatz \cite{Cerezo2021}, and quantum machine learning (QML) \cite{Biamonte2017} may be used to generate wavefunctions which encode the maximum independent set with an efficacy that may outperform classical computers. Indeed, a demonstration at Harvard in collaboration with QuEra shows that neutral-atom quantum processors can solve the MIS problem with a superlinear quantum speedup compared to a class of generic classical algorithms~\cite{ebadi2022quantum}. 

\newpage

%
%
%
%
%
%
\begin{table}[!ht]\label{fig:molecular_docking}
    \centering
    \begin{tabular}{|c|}
    \hline
         \parbox{\linewidth}{
         \vspace{-3mm}\section[Maximum clique and an application to molecular docking]{Maximum Clique}\label{sec:Clique2}\vspace{-3mm}
         \justify{\textbf{\LARGE{
\texttt{Example}:\quad Molecular docking\\
\texttt{Solution}:\quad Quantum ground state encoding}}
}}

\\\quad \\
         \hline
    \includegraphics[width=1\textwidth]{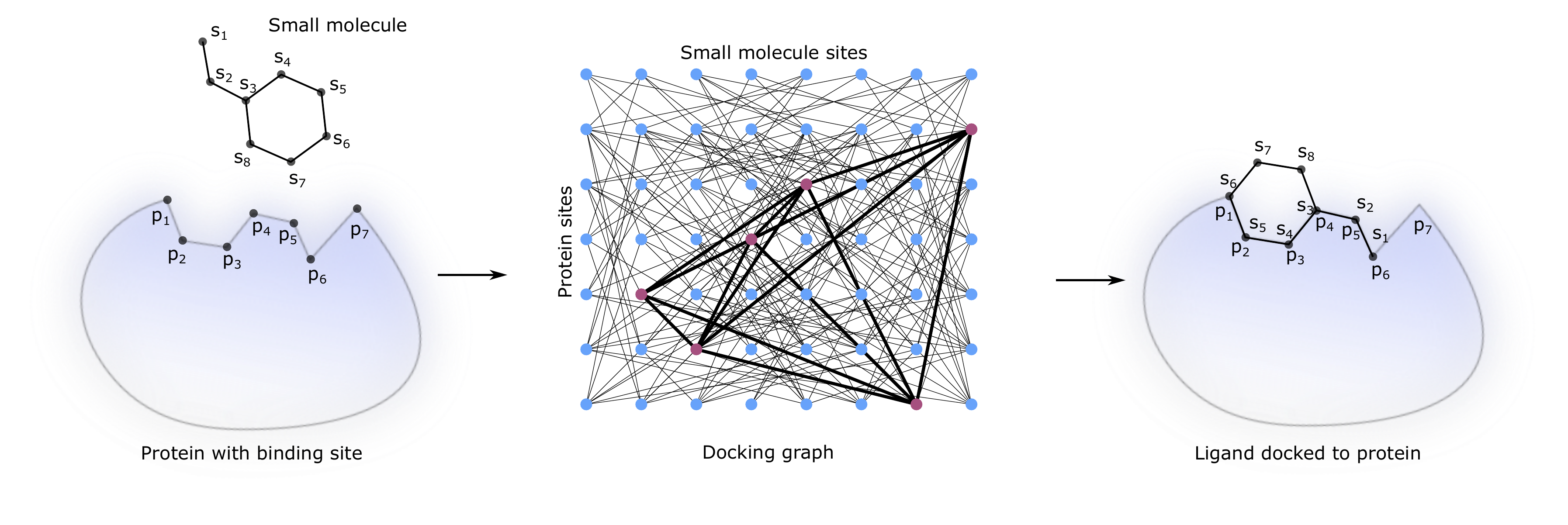}\\
    \parbox{0.9\linewidth}{ \justify{
    \textit{\textbf{Top:} An example of a molecular docking optimization problem. Left: A small molecule (``ligand") and a protein not bound to each other. The challenge is to find the correct docking pose of the ligand to the protein, which if it binds strongly enough, can inhibit the protein to cure diseases such as cancer. Center: The docking graph corresponding to the protein-ligand pair. Each vertex is a tuple $(p,s)$ matching a point $p$ on the surface of the protein to a point $s$ on the shape of the ligand. Each edge corresponds to a geometric match (successful superposition) of two ligand-protein atom pairs based to their shape. The maximum clique corresponds to the maximum valid matching of the ligand docking to the protein. Right: Ligand bound to the protein in the docking pose given by the maximum clique.
    }
    
    \quad
    
    A \emph{maximum clique} of a graph is the largest subset of vertices that are fully connected \cite{bomze1999maximum}. This means a clique forms a complete subgraph, and every vertex is adjacent to every other vertex in the clique. The maximum clique is the natural complement to MIS. A complement graph $\overline G$ is the graph that includes every vertex in $G$, and every edge between vertices that is \emph{not} in $G$. Because no two vertices in an independent set are adjacent, every vertex in the independent set is connected in the complement graph, forming a clique. Thus, the MIS vertex set $I$ of a graph $G$ is equivalent to the maximum clique of the complement graph $\overline G$, and vice versa. \\
    
    }} \\
    \hline
    \end{tabular}
\end{table}

Proteins play a key role in almost any disease \cite{alberts2022molecular}. To cure a disease, one typically needs to identify other molecules (ligands) that bind to a specific site of the target protein of interest to modulate the function of the protein (e.g. to inhibit its function). There exist different types of ligands, such as small molecules or other proteins. The most common type of ligands are small-molecule ligands, which have the advantage that they can typically be administered orally via tablets, are easy to distribute, and are inexpensive to produce. 

The information on how small molecules bind to proteins is extremely valuable in the fields of molecular biology and drug discovery. In the field of drug discovery, the binding mode of a ligand allows us to understand how the ligand exerts its function and enables optimization of the ligand based on the available structural information. Algorithms that can predict how a ligand binds to a protein are called docking methods \cite{gschwend1996molecular}. Docking methods consist typically of two components: a search algorithm that explores the conformational space of the ligand and the protein, and a scoring function that assigns each generated conformation a score. The final docking pose is the one that has obtained the best docking score. Docking methods are also used in structure-based virtual screenings, in which a large ligand library is screened against a given target protein to identify the ligands that will bind to it \cite{gorgulla2020open}. A large number of docking methods have been developed over the past decades, which deploy a large number of strategies, from gradient-based methods \cite{Flachsenberg2020} to swarm intelligence-based approaches \cite{eberhart95}. 

Combinatorial approaches to molecular docking have been explored early on in the history of molecular docking. Here we describe the one outlined in Ref.~\cite{kuhl_combinatorial_1984}, in which the docking problem (DP) is defined by two sets of points, $P=\{p_1,...,p_m\}$ and $S=\{s_1,...,s_n\}$ in $\mathbb{R}^3$. The points in $P$ define the surface of the protein, and the points in $S$ describe the shape of the small molecule (also called ligand in this context). The points can represent different features of the protein or ligand, such as atoms or function groups. Both the protein and the ligand are modeled to be rigid, and even though proteins and ligands are often flexible, their minimum energy conformation can for instance be used to represent a single shape of these molecules. A matching between the protein and the ligand is a set of pairs in $P\times S$ which can be superimposed to each other simultaneously, and the goal is to find a maximal matching. While unconstrained maximum matching is efficiently solvable using linear programming, the docking problem has edge constraints given by the shape of the ligand and protein, which makes the problem hard. Thus, the docking problem can be restated in combinatorial form, which is a special case of the well-known assignment problem \cite{SCHRIJVER20051} with pair constraints. Any assignment problem with pair constraints can be transformed into an independent set problem, which is in general NP-complete. The docking problem imposes a large number of geometric constraints on the assignment problem, by constraining pairs of matchings to be satisfiable geometrically. Within this approach, one can define distance matrices $D_{pp'}$ and $D_{ss'}$, which contain the pairwise distances between protein points and ligand points, respectively. These can be used to create a docking graph $G$ by setting an edge between pairs $(p,s)$ and $(p',s')$ if and only if the difference in distances is less than some tolerance $|D_{pp'} - D_{ss'}| < \epsilon$. Every maximal clique of $G$ corresponds to a maximal matching between protein points $P$ and ligand points $S$. 

The above approach to finding a maximum matching of a protein-ligand pair can be extended in multiple ways. One way is to not only consider whether the shape of a protein-ligand pair matches, but also whether matching atoms in the molecule have favorable or unfavorable interactions \cite{kuhl_combinatorial_1984}. The problem remains a maximum clique problem and thus can be solved by the Rydberg quantum simulator. Another way to generalize the method is to use weighted graphs to allow for more fine-grained modeling of the protein-ligand interactions, leading to a maximum weighted clique problem.

\subsection*{Quantum implementations}
There are many ways to compute the maximum clique of a graph using neutral-atom quantum computers, by finding the maximum independent set of the complement graph. The most natural implementation is adiabatic quantum computing \cite{Albash2018}, where the state of a system is slowly evolved to follow the ground state of some time-dependent Hamiltonian. The ground state of the Hamiltonian may be chosen to encode the maximum clique of a graph by reducing the problem to maximum weight independent set on unit disk graphs \cite{nguyen2023}. In this way, the measurement of the final wavefunction of the time-evolved system encodes the solution to the protein docking problem. Beyond adiabatic protocols, variational protocols such as the QAOA \cite{farhi2014quantum}, general variational ansatz \cite{Cerezo2021}, Gaussian Boson sampling \cite{banchi_molecular_2020}, and quantum machine learning (QML) \cite{Biamonte2017} may be used to generate wavefunctions which encode solutions with an efficacy that may outperform classical computers. Indeed, a demonstration at Harvard in collaboration with QuEra shows that neutral-atom quantum processors can solve the MIS problem with a superlinear quantum speedup compared to a class of generic classical algorithms~\cite{ebadi2022quantum}. 

\newpage

%
%
%
%
%
%

\begin{table}[h]\label{network_immunization}
\begin{tabular}{|cl|}
\hline
\multicolumn{2}{|c|}{\parbox{\linewidth}{
\vspace{-3mm}\section[Minimum vertex cover and an application to network immunization]{Minimum vertex cover}\label{sec:vertex-cover}\vspace{-3mm}
\justify{\textbf{\LARGE{
\texttt{Example}:\quad Network immunization\\
\texttt{Solution}:\quad Quantum ground state encoding}}
}}}\\\quad & \quad\\ \hline
\parbox{0.4\linewidth}{\includegraphics[scale=2]{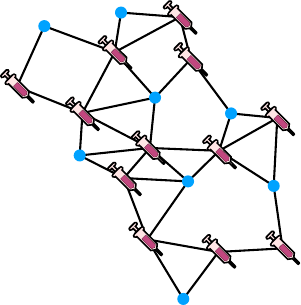}}

&
\parbox{0.565\linewidth}{\justify{
\textit{\textbf{Left:} An example network immunization problem. Objects represented as vertices, such as computers, hospitals, people, etc., may be infected by adjacent vertices connected by an edge. The challenge is to choose some subset of vertices to \textbf{immunize} to reduce the spread of infection. A vertex cover removes all edges from the network and thus perfectly stops the spread of infection; a minimum vertex cover perfectly stops the infection with the minimum resources.}

\quad

A \emph{vertex cover} $V'$ of a graph $G$ is some subset of vertices $V'\subseteq V$ such that each edge in $G$ is adjacent to at least one vertex in the vertex cover set. Naturally, the set of all vertices forms a vertex cover, and the minimum vertex cover chooses the smallest set of vertices. Usefully, given an independent set $I_m$, the complement vertices $V\backslash I_m$ is a vertex cover. For this reason, the maximum independent set and minimum vertex cover are complementary problems: The MIS $I$ of a graph $G$ is equivalently the minimum vertex cover $V'=V\backslash I$, and vice versa. As each edge has at least one vertex in the vertex cover, the graph $G\backslash V'$ has no edges and is equivalently an independent set. Similarly, the weighted versions of the maximum independent set and minimum vertex cover problems are complementary to each other as well. 

}}
\\\quad&\quad\\\hline

\end{tabular}
\end{table}

Network immunization is one of the interesting applications of the vertex cover problem~\cite{muhammad2017}. Here, the goal is to reduce the spread of some virus or other malevolent entity in a network \cite{eubank2004modelling}. For example, the entity could be a computer virus on a local area network, power fluctuations in a power grid, disinformation spreading on social media, or a pathogen spreading between hospitals. The spread of this entity can be reduced by selectively \emph{immunizing} a selection of nodes, which reduce the possible pathways for the entity to spread. The goal is to then intelligently select nodes so that 
the spread is completely cut, while the resources required are minimized.

The spread of the virus can be modeled by a graph $G$, where vertices represent possible targets for the virus to infect. An edge between vertices represents the possibility for either vertex to infect the other. Each edge may be weighted $w_{ij}$ by e.g.~the probability for vertex $i$ to infect vertex $j$. Usually, the dynamics of an infection network are modeled in one of two ways~\cite{Pastor-Satorras2015Epidemic}. The first is SIS or susceptible-infected-susceptible, where each vertex may become infected, recover, and become infected again. The alternate model is SIR, or susceptible-infected-recovered, where after infection the vertex is immune to reinfection (or dead). The spread is modeled by assuming that at some initial time, some (random or maliciously chosen) vertices are infected, and then the virus spreads through the network in a diffusive manner. The rate of spread can be characterized by the largest eigenvalue of the Laplacian of the graph, which equivalently is the fastest diffusive time scale in the graph.

The spread of the virus can be slowed by choosing some vertices to \emph{immunize}. An immunized vertex cannot be infected by the virus and thus cannot spread the virus to its neighbors. However, resources may be limited, and thus some optimal subset must be chosen to minimize spread. The $k$-vertex immunization problem is a specific version of this. Given $k$ vertices to choose to immunize, choose $n\leq k$ vertices to minimize the rate of spread on the graph. For example, one strategy may be to choose vertices that have high connectivity, which minimizes the infectious potential those vertices may have if infected.

A perfect immunization strategy would be to choose a subset of vertices that forms a vertex cover of the graph. Because each edge is incident to at least one vertex in the set of the vertex cover, there will be no remaining edges over which the virus can spread, and the virus is immediately and perfectly contained. However, because of the expense of immunizing many nodes, it is optimal to choose the minimum vertex cover of the graph, which perfectly contains the spread with the minimum number of immunizations. An imperfect immunization strategy would occur if $k$ is less than the minimum vertex cover size. In this case, the challenge would be to choose $k$ of $N$ vertices such that the largest eigenvalue of the reduced subgraph is minimal, which is the choice that minimizes the spread on the graph. In this case, solutions to $k$-vertex immunization are not vertex covers. One may also consider the case where each vertex has some weights, which can represent the cost or other utilities; this translates to the minimum weight vertex cover problem. 

\quad

In general, minimum vertex cover may be useful for problems where network resiliency is an important consideration. For example, analyzing the vertex covers of a supply chain may inform where to add redundant suppliers to reduce supply chain disruptions, or investigating the network topology of a secure local area computer network may inform strategies that reduce disruptions under cyberattacks. Minimum vertex cover can also be used in other network-flow applications, such as defending against tax avoidance \cite{fisman2004tax} or half-duplex mesh networks.
\subsection*{Quantum implementations}
There are many ways to compute the minimum vertex cover of a graph using neutral-atom quantum computers, by computing the maximum independent set $I$ of the graph, and then using the complement $V'=V \backslash I$. The most natural implementation is adiabatic quantum computing \cite{Albash2018}, where the state of a system is slowly evolved to follow the ground state of some time-dependent Hamiltonian. The ground state of the Hamiltonian may be chosen to encode the maximum independent set of a graph in an extremely natural manner using the Rydberg blockade and detuning fields, and so measuring the final wavefunction of the time-evolved system encodes the maximum independent set of the problem. If the problem cannot be encoded into a unit disk graph as may be the case for non-geometric graphs, it can alternatively be reduced to a unit disk graph using extra ancillary qubits~\cite{nguyen2023}. Beyond adiabatic protocols, variational protocols such as the QAOA \cite{farhi2014quantum}, general variational ansatz \cite{Cerezo2021}, and quantum machine learning (QML) \cite{Biamonte2017} may be used to generate wavefunctions which encode the maximum independent set with an efficacy that may outperform classical computers. Indeed, a demonstration at Harvard in collaboration with QuEra shows that neutral-atom quantum processors can solve the MIS problem with a superlinear quantum speedup compared to a class of generic classical algorithms~\cite{ebadi2022quantum}. 

For network immunization problems where the resources $k$ are less than the minimum vertex cover size, a hybrid variational scheme can be used to compute good immunization strategies. In this case, the neutral-atom device can be used in a sampling mode, for example, by providing independent sets from a variational ansatz, which can be used to initialize a classical post-processing step that removes vertices from the quantum-provided solution until the resource requirements are met. The objective function to variationally minimize is then the rate of spread on the post-processed solution. Because the classical post-processing step is biased by initializing its optimization from the samples of the neutral-atom quantum computer, it may gain an advantage over a purely classical implementation.

\newpage

%
%
%
%
%
%

\begin{table}[h]\label{minimum_dominating_set}
\begin{tabular}{|cl|}
\hline
\multicolumn{2}{|c|}{\parbox{\linewidth}{
\vspace{-3mm}\section[Minimum dominating set and an application to 5G network optimization]{Minimum dominating set}\label{sec:dominating-set}\vspace{-3mm}
\justify{\textbf{\LARGE{
\texttt{Example}:\quad Device-to-device 5G networks\\
\texttt{Solution}:\quad Quantum sampling \& QML}}
}}}
\\\quad & \quad\\ \hline

\parbox{0.4\linewidth}{\includegraphics[scale=2]{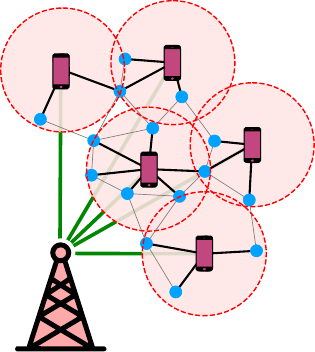}
}

&
\parbox{0.575\linewidth}{\justify{
\textit{\textbf{Left:} An example of a minimum dominating set for 5G ad-hoc wireless coverage. A distribution of cell phones or antennas in a geographic area (blue) wish to connect to the internet but may overload available bandwidth if every object gets connected to the nearby cell tower (green). Instead, a subset of cell phones can act as gateways (red), connecting to the cell network and providing internet access through Bluetooth or Wi-Fi to neighboring nodes. To minimize telecommunication bandwidth, it is preferable to minimize the number of connections to the cell tower. Requiring every node to be connected to the internet is a dominating set constraint, while minimizing the number of cell tower connections demands a minimum dominating set arrangement.}

\quad

Formally, a \emph{dominating set} is a subset of vertices $D\subseteq V$ of a graph $G$ that has the property that every vertex in $G$ is either in $D$, or adjacent to a vertex in $D$. Naturally, the set of all vertices forms a dominating set, and the minimum dominating set (MDS) chooses the smallest set of such vertices. Usefully, every maximal independent set is also a dominating set (but not the converse, as dominating sets may violate the independence constraint). This is simple to show by a proof by contrapositive: if some independent set $I$ is not a dominating set, there is some vertex $i$ that is not adjacent to $I$; so $I\cup\{i\}$ is also an independent set; thus $I$ is not a maximal independent set, as more vertices can be added. By contrapositive, maximal independent sets are dominating sets; in fact, every maximal independent set is also a minimal dominating set, but may not be the minimum. The minimum maximal independent set is the smallest maximal independent set; it is also called the minimum independent dominating set, which is generally a very good proxy for the minimum dominating set (without the independence constraint).
}}

\\\quad&\quad\\\hline

\end{tabular}
\end{table}

\quad

One example application where the minimum dominating set problem is useful is antenna placement that guarantees coverage for all users \cite{Gianluca2014, Monserrat2015}. Suppose there is some set of wireless nodes (computers, cell phones, distributed sensors, etc.) that have spatial location information $\{\vec x_i\}$. For simplicity, suppose each node can connect to every other node within a range $r$. This is equivalently a unit disk graph $G$ with vertices as the set of wireless nodes, and edges as the set of connected links. To connect each node to a network, one can choose some subset of locations to serve as \emph{gateway} nodes that are connected to the wider internet and serve as a relay for all nodes within range. Every node within range of a gateway node can connect to the broader internet through a wireless connection to the gateway node, which serves to forward traffic. Given the requirement that every node in the network be within range of at least one gateway node and so be connected to the internet, the gateway nodes serve as a dominating set over the graph $G$. If one avoids direct communication between the gateway nodes, it forms an independent dominating set.

As a particular example of antenna placement and dominating sets, consider a device-to-device (D2D) communication network between cell phones or network-connected cars, which is part of a 5G architecture \cite{Monserrat2015}. In this example, there are many individual nodes (for example, a crowd at a concert or dense autonomous traffic on a highway). Each node may communicate with other nodes within range (the ``User", or ``U" plane) using short-range protocols, such as Bluetooth, ad hoc Wi-Fi, or near-field communication (NFC). Alternatively, each node may communicate with the broader wireless network of cell towers (the ``Control", or ``C" plane). Each node may communicate with nearby nodes (for example, position coordination between adjacent autonomous cars), or with the broader internet (for example, sending pictures of a concert to a friend). However, connecting every node to the broader network of cell towers may overload bandwidth limitations. To circumvent this problem, one can choose a subset of nodes to serve as the gateway nodes, connect with the cell network, and provide internet access to nearby nodes. Every other node in the area may connect to the broader internet through a direct connection to an in-range gateway node, reducing the cell tower bandwidth needs.

Requiring that every node needs to be a gateway, or within range of a gateway, naturally forms a dominating set constraint on the set of gateway nodes. It is natural to desire small dominating sets, which minimize the number of connections to the cell network and thus minimize bandwidth. Given a weight $w_i$ for each node (for example, the battery or connection quality for each cell phone), the minimum weighted dominating set is the optimal choice of gateway nodes, which minimizes the bandwidth resources. Again, if the gateway nodes are not in range of each other directly, one has the minimum independent weighted dominating set. 

\quad

In general, minimum dominating sets may be useful when there are networks and relationships that require a hierarchical structure. For example, a dominating set could be used to optimize supply chain logistics by concentrating stock to a subset of warehouses or choosing leader-follower assignments for a distributed computing cluster. Anywhere where one may need to minimize resource use while keeping every node connected will be a good use for minimum dominating sets.

\subsection*{Quantum implementations}

While maximum independent sets and related problems are naturally mapped onto the ground state of a Rydberg Hamiltonian, this is not the case for dominating sets. This is because good dominating sets are minimum maximal independent sets, a difficult problem of both minimization and maximization. Additionally, minimum dominating sets may violate the independent set constraints in certain situations. For this reason, a quantum application must work in a variational hybrid mode or using a quantum machine learning approach, which includes both quantum and classical resources. Here, the neutral-atom device can, for example, work in a sampling mode, providing independent sets from some weighted probability distribution determined by the variational wavefunction $|\psi(\vec\alpha)\rangle$. In this variational mode, the state may even weakly violate the independent set constraint by choosing particular time-dependent Hamiltonians. Then, each set sampled from the machine is used as the initial point for a classical post-processing step, which adds vertices until the solution is a dominating set. The objective function, which is variationally minimized in a hybrid quantum-classical loop, is the average size of the dominating sets provided from the classical post-processing step sampled from the variational wavefunction. Because the classical post-processing step is biased by initializing its optimization from independent sets sampled from a quantum probability distribution of the neutral-atom quantum computer, it may gain an advantage over a purely classical implementation.

\newpage

%
%
%
%
%
%

\begin{table}[h]\label{ad_hoc_networks}
\begin{tabular}{|cl|}
\hline
\multicolumn{2}{|c|}{\parbox{\linewidth}{
\vspace{-3mm} \section[Minimum connected dominating set and an application to ad-hoc networks]{Minimum connected dominating set}\label{sec:connected-dominating-set} \vspace{-3mm}
\justify{\textbf{\LARGE{\
\texttt{Example}:\quad Ad-hoc network routing\\
\texttt{Solution}:\quad Quantum sampling \& QML}}
}}}
\\\quad & \quad\\ \hline
\quad&\quad\\
\hspace{-20mm} \parbox{0.45\linewidth}{\includegraphics[scale=2]{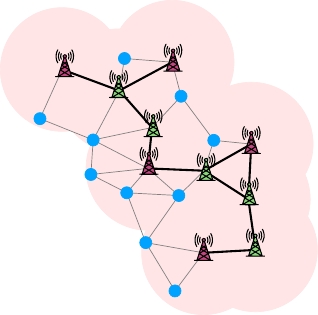}}
&
\hspace{-25mm}
\parbox{0.45\linewidth}{
\justify{\textit{\textbf{Left:} An example connected dominating set for ad-hoc network routing. A distribution of wireless nodes can connect with every nearby node, forming a unit disk graph. Occasionally, nodes may communicate by sending packets to the network, which are forwarded from node to node according to some routing protocol. Under a proactive dominating set based routing protocol, the dominating nodes (red) form gateways that provide the routing instructions between nodes upon request, reducing network upkeep and increasing speed and reliability. A minimum connected dominating set minimizes the number of gateway nodes, reducing network overhead.}}
}
\\
\qquad\parbox{0.6\linewidth}{\parbox{1.55\linewidth}{
\justify{
A \emph{connected dominating set} $D_c$ is a dominating set with the extra property that the subgraph induced by $D_c$ is connected, i.e., there is a path between each vertex using only vertices in $D_c$. A minimum connected dominating set (MCDS) is the smallest such set. Unfortunately, the connectivity requirement violates the independence constraint, so the MCDS problem does not directly fall into the class of independent set problems. However, given an MDS, it is possible to find a connected dominating set within a factor of three of the MCDS. Each vertex in the MDS is at most a distance of three from another MDS vertex, and so one can add at most two extra vertices per vertex to create some connected graph. In this way, $3|\texttt{MDS}|\geq |\texttt{MCDS}|$. Additionally, since an MCDS is also a dominating set by definition, $|\texttt{MCDS}|\geq |\texttt{MDS}|$. Therefore, by adding vertices to an MDS to connect it, it is possible to generate a small (but may not minimum) connected dominating set within a factor of three of the minimum connected dominating set.

On unit disk graphs, there is also a bound between the size of a maximum independent set and a minimum connected dominating set. For example, Ref.~\cite{wu2006minimum} proves that $\text{mis}(G) \leq 3.8 * \text{cds}(G) + 1.2$, where $\text{mis}(G)$ is the size of a maximum independent set and $\text{cds}(G)$ is the size of a MCDS in the graph $G$. Constructing a maximal independent set and then connecting it by adding more vertices is often a good heuristic method to find the MCDS approximately.

}}}&
\\\quad&\quad\\\hline

\end{tabular}
\end{table}

One example application of the connected dominating set problem is ad-hoc networks \cite{wu2006minimum}. An ad-hoc network is a set of nodes that are connected and communicate wirelessly by links. For example, these links could be distributed to sensors in a city, autonomous cars on a highway, soldiers on a battlefield, and many more. Every node in the network can forward data between connected nodes, and so forms a local area network where every node can, in principle, communicate and coordinate with every other. This network architecture has the advantage of being decentralized, in the sense that there is no reliance on pre-existing infrastructure and must be self-configuring and dynamic. However, instead of communicating through a centralized infrastructure, an ad-hoc network relies on intermediate nodes forwarding data, which means that communication may suffer from longer latency or poorer connectivity if the routing is ill-configured: an excess of data may overload the forwarding capability of a node or an ill-optimized path between nodes may increase the time to transmit data. It is thus imperative to optimize routing protocols that utilize the network efficiently while being robust and dynamic to changes in the connectivity.

One application of connected dominating sets to ad-hoc networks is for routing protocols, which is the method by which data is transmitted across the ad-hoc network. The simplest and most na\"ive protocol is \emph{flooding}. Under the flooding protocol, each message is sent to every other node in the network. This is done by instructing every node to forward every message that it has not previously received to every other node. In this way, the message is ``flooded" across the network until every node, including the destination node, has received it. However, such a method suffers from excessive overhead and poor scaling as the number of nodes increases: each message gets seen by every node, and every node may periodically send messages.

A more efficient protocol is \emph{proactive routing}, where data is routed along the shortest path between nodes. In this implementation, the route between each node is pre-computed (for example, using Dijkstra's algorithm), and then data is distributed to every node in the network. Each node holds a table of every other node in the network, which gives the choice of which adjacent node to pass each packet to. In this way, each packet of data is passed in a single efficient path between source and destination nodes. However, this implementation suffers from some initialization overhead. Whenever the network topology changes (from sensors going offline, cars moving out of range, and so forth), the path information must be recalculated, and routing information be redistributed to every node in the network. This overhead may get especially excessive if the ad-hoc network is extremely dynamic, as might be the case for nodes that are quickly moving around spatially or often disconnecting.

A more efficient proactive routing protocol, which partially circumvents the recalculation step, is the \emph{dominating-set-based routing}. With this protocol, each node does not need to hold or calculate the routing table, which avoids the costly recalculation step. Instead, a small dominating set of nodes is chosen as gateway nodes, which hold the routing tables of paths between each node that it dominates, and every other node. A node may transmit a message to another node in the following manner. First, a node generates its message and identifies the receiving node. Then, the node queries the gateway node that is its dominator for routing information between itself and the receiving node. Because these gateway nodes form a dominating set, this is possible for every node in the network. Then, the gateway provides a pathway from its precomputed table to the querying node, which is then routed on the ad-hoc network over which each node forwards to the next. Finally, the message and routing pathway are packaged together and passed from node to node along the network. Each node along the route simply looks up the next node in the route and passes it accordingly.

Note that this dominating-set-based routing protocol does not require a connected dominating set. The connected condition adds resilience to the system: if a connection between the nodes breaks (for example, from a node going offline or moving out of range), the original path provided by the gateway nodes will fail. In this case, the message can be rerouted along the components of the connected dominating set, which forms a communication backbone. Alternatively, if traffic is light, the connected dominating set backbone could be used to pass every message, simplifying what might otherwise be an extremely dynamic connection graph and assuring that messages are received with a higher probability.

\quad

In general, MCDS may be useful when objects need to be passed around some network. The ad-hoc routing example above covers an extremely large number of cases in telecommunications for moving information around a dynamically changing network. However, the MCDS may be useful for more general network analysis. For example, finding connected dominating sets in the epidemiological spread graph may be useful to identify routes where a virus or pathogen may quickly spread through the network.

\subsection*{Quantum implementations}

While maximum independent set and related problems are naturally mapped onto the ground state of a Rydberg Hamiltonian, this is not the case for connected dominating sets. This is because good connected dominating sets are, by definition, not independent sets, although they are closely related to dominating sets and independent sets. For this reason, a quantum application must work in a variational hybrid mode or using a quantum machine learning approach, which includes both quantum and classical resources. Here, the neutral-atom quantum computer can, for example, work in a sampling mode, providing independent sets from some weighted probability distribution determined by the variational wavefunction $|\psi(\vec\alpha)\rangle$. Then, each independent set sampled from the machine is used as the initial point for a classical post-processing step, which adds vertices until the solution is a connected dominating set. One implementation of this would be to first construct a maximal independent set (e.g.~a minimal dominating set) from the quantum-initialized independent set, and then add vertices to make the dominating set connected. The objective function, which is variationally minimized in a hybrid quantum-classical loop, is the average size of the connected dominating sets provided from the classical post-processing step sampled from the variational wavefunction. Because the classical post-processing step is biased by initializing its optimization from independent sets sampled from a quantum probability distribution of the neutral-atom quantum computer, it may gain an advantage over a purely classical implementation.

\newpage

%
%
%
%
%
%

\begin{table}[h]\label{loss_networks}
\begin{tabular}{|cl|}
\hline
\multicolumn{2}{|c|}{\parbox{\linewidth}{
\vspace{-3mm}\section[Independent set sampling and an application to telecommunication loss networks]{Counting Independent sets}\label{sec:counting-IS}\vspace{-3mm}
\justify{\textbf{\LARGE{
\texttt{Example}:\quad Telecommunication loss networks\\
\texttt{Solution}:\quad Quantum sampling}}
}}}\\\quad & \quad\\
\hline
\parbox{0.55\linewidth}{\includegraphics[scale=2]{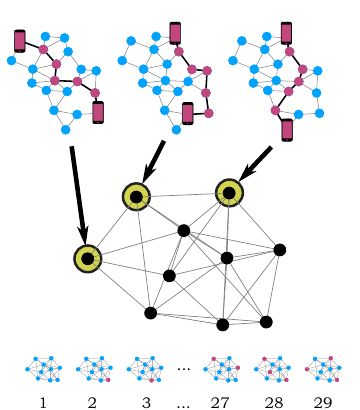}
\parbox{0.9\linewidth}{\justify{\textit{
\textbf{Top:} An example route interaction graph. Calls between customers (top) are routed between each other (red) on some network graph. No two calls can share a link (black edges). To compute the failure to complete a call, one must compute statistics on the route interaction graph (middle) by counting independent sets. Each route is a vertex, and there is an edge if two routes share at least one link. For example, the left and right routes share no links and thus can both be simultaneously completed, while the left and center routes share a link and thus routing must be delayed or fail, reducing the performance of the network. Allowed configurations of calls are independent sets in this route interaction graph. Analyzing probability distributions of independent sets is useful for understanding the capacity and performance of the routing network.}
}}}

&
\parbox{0.425\linewidth}{\justify{
Beyond finding large independent sets, another interesting task is to \textit{count} the number of independent sets. Unlike its \texttt{NP-complete} decision counterpart (``there exists an independent set of size $k$"), or \texttt{NP-hard} combinatoric counterpart (``here is an independent set of size $k$"), computing the number of independent sets is a \texttt{\#P-complete} (pronounced sharp P complete) counting problem \cite{Bandyopadhyay2008}. These \texttt{\#P-complete} problems are known to be at least as hard as \texttt{NP-complete} problems. A fundamental IS counting problem is to compute the independent set \emph{partition function}
\begin{equation*}
    \mathcal Z = \sum_{I\in \{ \text{IS}(G) \}}P(I)
\end{equation*}
given some weighting function $P(I)\in[0,1]$, where $I$ is an independent set. Given an interpretation of $P$ as an unnormalized probability, the partition function $\mathcal Z$ acts as the normalization for computing expectation values and is widely used in statistical physics. If $P=1$ for all independent sets, the partition function simply counts the number of independent sets of a graph. For more general weights, the partition function would count independent sets weighted by $P(I)$. For this reason, computing the partition function is \texttt{\#P-complete}. Similarly, computing average values of a function over a distribution of independent sets is also \texttt{\#P-complete}. Given some function $F(I)\to\mathbbm R$, an average is
\begin{equation*}
    \langle F\rangle\;=\; \frac{1}{\mathcal Z}\sum_{I\in \{ \text{IS}(G) \} }P(I)F(I).
\end{equation*}
Since the partition function is hard to calculate, computing expectation values is typically hard as well.
}}
\\\quad&\quad\\\hline

\end{tabular}
\end{table}

\quad

One example application where counting independent sets is useful is loss networks in telecommunications \cite{LOUTH199445,Kelly1991}. Suppose there is some network of nodes that are connected by \emph{links}. For this example, consider an early telephone network, where each vertex is a user that can make or forward a call. Occasionally, a user at node $A$ may wish to make a call with a user at node $B$. These users are connected by a \emph{route} between $A$ and $B$, which makes exclusive use of the links between every node on the route. More likely, many users at different nodes may wish to connect with each other through many routes. However, because each link is used exclusively by a single route, not every route is possible at the same time. For example, if a new call is made whose route uses a link that is already in use by another call, it will fail to connect, reducing the capacity of the network. To construct a reliable network, it is thus imperative to determine the conditions where the network may fail, to best utilize resources and build robustness in the system.

One important metric to optimize in a phone network is to determine the capacity of the network, and how often calls fail to connect due to an excess of links being used. If there are many concurrent calls on the network, the route of a new call may use a link that is in use by another call and will fail to connect. Such a failure to connect a call may have real-life consequences, as a high failure rate will reduce the reliability of the network and the satisfaction of customers. It is then imperative to design routes and networks that have a high probability of new calls being able to connect, even if the network is heavily used.

The exclusivity of no two routes sharing links naturally forms an independence constraint on the possible simultaneous routes. Suppose a network graph with a set of routes $\{r\}$, where each route $R_r$ uses some subset of links of the network graph. This set of routes defines a \emph{route interaction graph} with an independence constraint. Each vertex is a route, and there is an edge between two vertices if the two routes share a link. A valid set of routes between multiple users forms an independent set on the route interaction graph, as no two routes share a link on the network. The size of the maximum independent set is thus the largest number of possible simultaneous phone calls on the network.

One simple model \cite{Kelly1991} which is used to calculate this failure probability is to suppose a steady state of calls being placed between customers. Here, each route is active with some small and independent probability $\nu$. As such, the probability of an independent set of size $|I|$ is $P(I)=\nu^{|I|}$ and the partition function is $\mathcal Z(G)=\sum_{I\in \{ \text{IS}(G) \} } \nu^{|I|}$. It can be shown that given this route density and choice of routes, the success probability of a call on route $r$ to complete is
\begin{equation*}
    \text{Probability of success on route }r\;=\;P(r)\;=\;\frac{\mathcal Z(G\backslash r)}{\mathcal Z(G)}
\end{equation*}
where $G\backslash r$ is the graph minus the route $r$ and $\mathcal Z$ is the partition function. Using this value, one can optimize the network by choosing good routes between each caller. Each choice of routes generates a different route interaction graph, which has some different failure probability per route $P(r)$. Some choices of routes have a minimum failure probability and so may be optimized to increase the reliability of the network under load. However, because computing this success probability is hard, a route optimization on a classical computer is similarly hard.

This simplistic example can, of course, be generalized and extended. For example, each link may have the capacity to route multiple calls at once, which increases the throughput and reliability of the network. Similarly, the probability distribution may be modified (for example, to include customers trying to call again until their call succeeds). Additionally, a call may take multiple routes, with some routes being optimal given some state of the network.

\quad

There are many other applications where counting independent sets and computing statistics of weighted IS distributions may be useful. Beyond call routing, these route interaction graphs can be used for scheduling problems, where agents (such as autonomous drones, airplanes, internet data, etc.) must route themselves between locations without collisions. In this sense, this loss network example extends far beyond telecommunications to many network analysis problems of this type. Generally, when making decisions under uncertainty where the state space has an independence constraint, being able to compute partition functions and expectation values may be important to inform optimal choices.

\subsection*{Quantum implementations}

While simulating distributions of independent sets may be a hard computational task classically, it is an extremely natural application for a neutral-atom quantum computer where the Rydberg blockade naturally encodes the independence constraint. Given some wavefunction $|\psi\rangle$ prepared using a variational algorithm, the probability of measuring an independent set $I$ is equal to $P(I) = |\langle I|\psi\rangle|^2$. This makes sampling independent sets from a biased distribution exceedingly natural if the wavefunction matches the probability weights. Furthermore, computing expectation values of functions is simple, as one only needs to average the function over many samples from the wavefunction. As an example wavefunction, suppose the weights form a Boltzmann distribution $P(I) = \exp(-\beta H(I))$. This is the case if calls on a phone network $P(I) = \nu^{|I|}=\exp(\log(\nu)|I|)$ if the Hamiltonian counts the independent size $H(I)=|I|$, and the probability takes the role of temperature $\beta = -\log(\nu)$. A quantum state can be constructed adiabatically or variationally~\cite{Wild2021} that could reproduce this probability distribution and thus may reproduce any classical statistics required.

In this mode, the quantum device acts as a \textit{variational quantum sampling machine} \cite{Amin2018}, where the wavefunction reproduces a probability distribution that is hard to calculate on a classical device. These sampling machines can be coupled with quantum machine learning algorithms \cite{gao2021enhancing} or classical post-processing to inform optimal choices to computationally intractable statistical problems.

\newpage

%
%
%
%
%
%

\begin{table}[h]\label{store_placement}
\begin{tabular}{|cl|}
\hline
\multicolumn{2}{|c|}{\parbox{\linewidth}{
\vspace{-3mm}\section[Maximal independent set and an application to incremental store placement]{Maximal independent set}\label{sec:maximal-IS}\vspace{-3mm}
\justify{\textbf{\LARGE{
\texttt{Example}:\quad Incremental store placement\\
\texttt{Solution}:\quad Quantum sampling \& QML}}
}}}
\\\quad & \quad\\ \hline
\parbox{0.4\linewidth}{\includegraphics[scale=2]{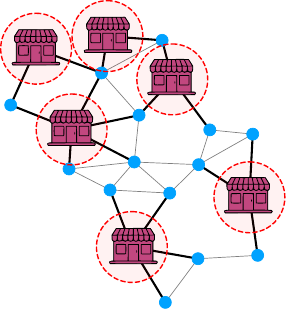}
}

&
\parbox{0.575\linewidth}{\justify{
\textit{\textbf{Left:} An example maximal independent set for store placement. Given a choice of store locations in a geographic area (blue) under the independence constraint that no two stores are within $d$ miles, a maximal independent set is a choice of store locations that blocks any new store locations from being built. In this example, each store is built one at a time, and store locations may become unavailable later. For this reason, simply choosing the MIS may be sub-optimal: as locations become unavailable, previously placed stores may accidentally block better solutions. Evaluating probability distributions of maximal independent sets is useful to inform which locations give the most flexibility in future store placement under uncertainty.}

\quad

A \emph{Maximal independent set} (mIS) $M$ is an independent set of a graph, which has the additional property that no vertex can be added to the set without violating the independence constraint. A maximum independent set is always a maximal independent set, but the opposite is not true. An mIS is also a \emph{dominating set}, which means, equivalently, that every vertex is either in $M$, or is adjacent to a vertex in $M$. For this reason, it is impossible to add extra elements to a maximal independent set, as every choice is ``blocked" by definition. It is easy to sample from the set of maximal independent sets using a greedy classical algorithm, by incrementally adding random vertices to a set under the independence constraint, until no further vertices can be added. However, as described in section \ref{sec:counting-IS}, sampling from a weighted distribution of independent sets, or counting the number of maximal independent sets, is hard.
}}
\\\quad&\quad\\\hline

\end{tabular}
\end{table}


One natural geometric example application where the independent set sampling is useful is in strategic planning, such as choosing locations for stores or buildings under constraint and uncertainty. Suppose there are some choices of store locations $\{\vec x_i\}$ that are subject to the constraint that no two buildings can be within $d$ miles of each other. This may be due to a zoning constraint, a heuristic business policy, or other restrictions. Each store, for example, reaches a certain number of customers depending on the local population density, or has a certain profit potential, which gives each store some weight value $w_i$. This naturally forms a unit disk graph with an independence constraint, where vertices are potential store locations, and edges represent the constraint that two store locations are too close together. The weight of an independent set $W_I$ is, for example, the number of customers reached or the total profit potential of the independent set choice $I$. The maximum weighted independent set is then the choice of stores which maximizes profit or the number of stores.

A similar problem of store placement entails incremental addition of stores and counting independent sets. While, in some cases, it may be possible to add many locations at once that optimize some sales metric, in many other cases it may be infeasible to add many locations at the exact same time. A more realistic version may be to add stores incrementally, where each location is added one at a time in a serial manner.

However, it may be unreasonable to expect that the choice of locations may stay the same in the future. For example, store locations may become unavailable, zoning regulations may change, or the weight representing the value of the store may change due to shifting population or infrastructure. For this reason, incrementally adding stores planned from an MIS solution may not be an optimal plan. If some number of planned future sites become unavailable, the already-built stores may accidentally block locations that would otherwise be part of a better independent set solution.

For this reason, computing statistics of weighted ensembles of maximal independent sets may be useful for incremental store placement. Instead of (suboptimally) rigidly choosing a maximum independent set and incrementally choosing locations, one can choose the next location incrementally, such that profit potential is maximized and future store locations can be chosen flexibly. Each maximal independent set can be considered a ``final plan" where one must stop building any more store locations, so that the set of maximal independent sets is the set of all \emph{site plans}. The weighted maximum independent set is, of course, the best choice, but there is some probability of failure which may make it sub-optimal under uncertainty. Each site plan $I_M$ has some probability of success $P(I_M)$ and site preference $w(I_M)$, which may be weighted by profit potential, customers served, and so forth. The expected site preference across all site plans is then the sum over all maximal independent sets, weighted by the site preference and success, normalized by the partition function:
\begin{equation*}
    \langle w\rangle = \frac{1}{\mathcal Z}\sum_{I_M \in\text{\{mIS\}}} w(I_M)P(I_M),\qquad\text{where}\qquad \mathcal Z =\sum_{I_M \in\text{ \{mIS\} }} P(I_M).
\end{equation*}
This expectation value can be modified to give a decision on the next store location by computing the average site preference given that store $s$ is in the plan, and then maximizing $\langle w\rangle_s$ over all possible choices:
\begin{equation*}
    \langle w\rangle_s = \frac{1}{\mathcal Z}\sum_{s\in I_M \in \{ \text{mIS} \} } w(I_M)P(I_M).
\end{equation*}
The best store location may not just have the largest utility, but is included in many maximal independent sets that are profitable and have a high chance of succeeding, has the largest value of $\langle w\rangle_s$, and thus is the best candidate for the next store. However, this requires taking a weighted average over all possible maximal independent sets, which is a \texttt{\#P-complete} counting problem.

\quad

There are many other applications of counting independent sets. For example, in the previous problem of routing calls in telecommunications, there may be multiple routes to connect two calls. Given a new call, it is optimal to choose a route that maximizes the future flexibility of the network to additional calls, which is in the same spirit as the incremental store placement problem. Additionally, counting the number of solutions to a scheduling problem may inform incremental task assignments which increase the flexibility of further choices in the face of future unknowns, or counting the number of maximal independent sets to a portfolio optimization problem informs the amount of anti-correlation in a market. Generally, if given some constrained problem for which a solution must be constructed in an incremental problem, counting maximal independent sets may be a way to choose strategies that maximize future flexibility and profit.

\subsection*{Quantum implementations}

While simulating distributions of independent sets may be a hard computational task classically, it is an extremely natural application for a neutral-atom quantum computer where the Rydberg blockade naturally encodes the independence constraint. Given some wavefunction $|\psi\rangle$ prepared using a variational algorithm, the probability of measuring an independent set $I$ is equal to $P(I) = |\langle I|\psi\rangle|^2$. In this way, the neutral-atom quantum computer can be used in a sampling mode to solve the incremental store placement problem using classical post-processing and quantum machine learning, in two steps. First, the neutral-atom device can be used to generate some variational wavefunction $|\psi(\vec\alpha)\rangle$, which can sample independent sets $\{I\}$. Given any of these quantum-sampled independent sets, a classical post-processing step can greedily add vertices to generate a distribution of maximal independent sets $\{I_M\}$, which serve as the ``site plans" for store placement. Using quantum machine learning to optimize the variational parameters $\vec \alpha$, the likelihood can be maximized to match the quantum-sampled probability distribution with the classical one $P_Q(I_M)\approx P(I_M)$. Then, the expectation value $\langle w\rangle_s$ can be measured using repeated sampling of the wavefunction and post-selection to compute the optimal next choice of store placement. Because the independent sets are sampled from a quantum wavefunction, the extra expressiveness may be able to capture classically hard probability distributions and gain a quantum advantage over classical-only implementations.

\newpage

%
%
%
%
%
%

\begin{table}[ht!]\label{task_scheduling}
\begin{tabular}{|cl|}
\hline
\multicolumn{2}{|c|}{
\parbox{\linewidth}{
\vspace{-4mm}
\section[Graph coloring and an application to the task scheduling problem]{Graph coloring}\label{sec:graph-coloring}
\vspace{-4mm}
\justify{\textbf{\LARGE{
\texttt{Example}:\quad Task Scheduling\\
\texttt{Solution}:\quad Quantum ground state encoding}}
}}}\\ \quad & \quad\\ \hline
\quad\parbox{0.5\linewidth}{\vspace{5mm}\includegraphics[scale=1.7]{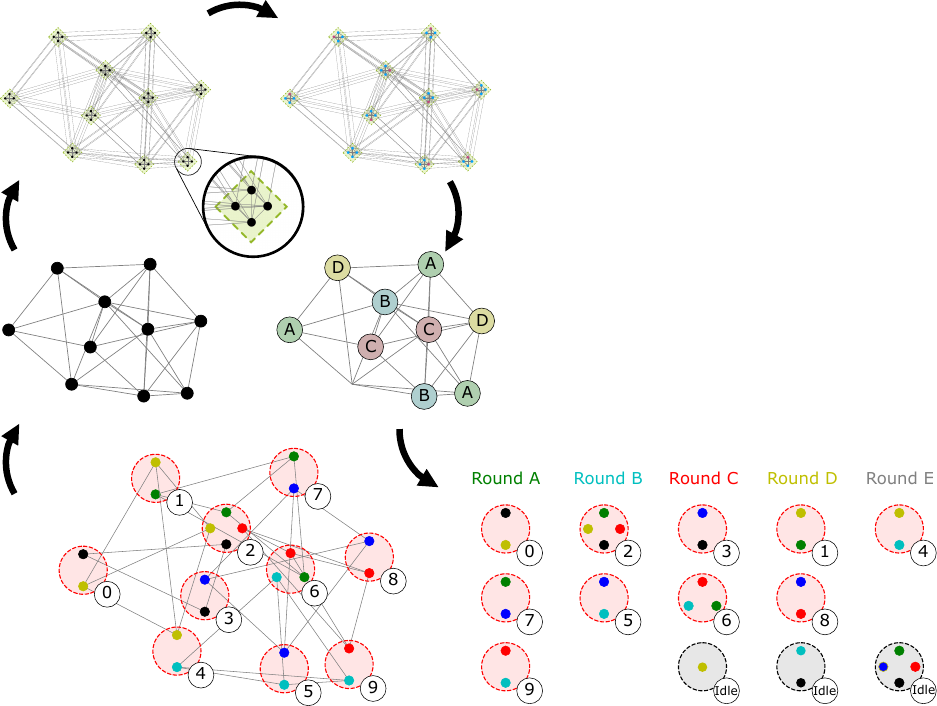}}

&
\parbox{0.45\linewidth}{

\justify{\textit{
\textbf{Left:} An example scheduling problem, graph coloring, and independent set mapping. A scheduling problem (bottom left) is made of tasks (pink dashed circles) accomplished by assets (dots in the pink dashed circles). In order to complete a task, it needs all the assets within the task. On the other hand, 
an asset can only be used in one task at a given time (round), giving an independence constraint on the graph of tasks, where there is an edge if two tasks share an asset (blue middle left). The goal is to find a schedule of tasks that maximize asset utilization by minimizing the number of rounds. This can be formulated as the $K$-coloring problem, where the number of colors, $K$, represents the number of rounds. 
This $K$-coloring problem can be reduced to an independent set problem (top) by finding the maximum independent set of a particular graph, which is shown here for $K=4$ colors. Solving the coloring problem finds a sequential schedule of tasks that best utilizes assets by minimizing the number of rounds (bottom). This task scheduling problem may be used for a diverse number of uses, from allocating machines to manufacturing widgets in a factory, to finding efficient routing schedules for autonomous vehicles subject to the constraint that no two routes (assets) overlap. \vspace{4.5cm}
}}
}
\vspace{-5mm}
\\
\multicolumn{2}{|c|}{
\parbox{0.95\linewidth}{\justify{

A closely related problem to the maximum independent set problem is the \emph{graph coloring} problem. Given $K$ ``colors" which may be assigned to each vertex of a graph, a graph coloring finds a choice of colors subject to the constraint that no vertices of the same color are adjacent. Finding the minimum number of colors, or the \emph{chromatic number} of the graph is \texttt{NP-complete} using a reduction from 3-SAT. The graph coloring is a natural extension of independent sets, as the vertices of each color of a coloring solution form an independent set.

Thus, graph $K$-coloring has an extremely natural reduction to the independent set problem, and it has a similar adjacency restriction. Given some graph $G$ of vertices $V$ and edges $E$, and a number of colors $K$, an alternative independent set graph $G'$ may be constructed by duplicating each vertex into $K$ vertices, one for each color. To enforce the condition of only one color per vertex, each of these $K$ vertices is fully connected to each other to form a clique. To enforce the condition that no two vertices of $G$ are the same color, $K$ edges are added between adjacent vertices connecting the requisite color vertices of each clique (top left). In this way, if a graph is $K$-colorable, the MIS of $G'$ has a cardinality $|V|$, corresponding to an MIS vertex in each clique labeling the color. If the graph is not $K$-colorable, the MIS of $G'$ has a cardinality $<|V|$. The cliques $V'$ with no MIS vertices (top right) correspond to the minimum number of vertices to remove from $G$ to make $G\backslash V$ exactly $K$-colorable, and cliques with an MIS vertex label the color.
}}}
\\\quad&\quad
\\\hline


\end{tabular}
\end{table}

One example application of independent sets and graph coloring is the scheduling problem. Here, the problem consists of some collection of \emph{tasks}, and some collection of \emph{assets} used to complete these tasks.

For example, these tasks could be manufacturing jobs on a factory floor, with the assets being the particular machine tooling required to complete the job. Alternatively, the tasks could be a set of sports games between different teams that may take place during a season, and the assets are the teams that join in each game. This naturally forms a general graph with an independence constraint, where vertices are tasks, and edges represent two tasks sharing one or more assets. In this way, an independent set is a choice of tasks in which each asset is used in at most one task. A maximum weighted independent set is the choice of tasks that maximizes some objective function over the tasks, such as the number of widgets produced in a factory or how time-critical it is to execute a particular task to restock a low inventory.

As a particular example, consider an example of a widget factory with widgets manufactured by different production lines (tasks). Each manufacturing line requires some subset of machines (assets), represented by colors in the graph above. Suppose the widget factory has some inventory of each kind of widget and is dynamically producing widgets so as to never run out of widgets in inventory as they are shipped out of the factory. The production is done in consecutive rounds, where in each round assets are assigned to a subset of tasks as an independent set. Between rounds, a new weighted maximum independent set is chosen based on the amount of inventory of each kind of widget: if the amount of stock is low, the task is given a higher weight, and if the stock is high, the task is given a lower weight. If the maximum weighted independent set is chosen for each round, on average a larger amount of manufacturing lines are being run each round, which maximizes the production capacity of the widget factory.

A more in-depth application is consecutive scheduling. Here, the goal is to complete every task in a minimum number of rounds. Between rounds, each asset is swapped between tasks, which incurs some cost. It is then profitable to minimize the number of rounds, which maximizes the utilization of assets. Given a choice of $K$ rounds, the allocation of which tasks to which round corresponds to a graph $K$-coloring. The color adjacency restriction is a natural extension of the independence constraint, as the vertices of each color form an independent set, and the one-and-only-one color choice naturally represents the restriction that each task be completed exactly once. The color of each vertex's solution then corresponds to the round that the task is allocated to.

As an example of consecutive scheduling, consider the same widget factory, except with a fixed schedule of rounds, where each manufacturing line is run exactly once. A $K$-coloring with $K$ minimized corresponds to the ordering of which manufacturing lines are turned on to produce widgets that minimize the number of rounds and thus maximize the production capacity of the widget factory. For the tasks and assets shown in the figure above, a 4-coloring does not color every vertex, and so the optimal four-round schedule excludes one manufacturing line.

\quad

In general, graph coloring may be useful as an extension of independent set problems, where one may need multiple complementary independent sets with small overlaps to maximize asset utilization or minimize scheduling overhead \cite{Gainanov2018}. In principle, the scheduling constraints can be generalized (for example, multiple indistinguishable assets) or have more complicated objective functions (for example, including the cost of moving a piece of machinery between lines on the factory floor). For example, this task scheduling may be useful for autonomous vehicle routing, where the role of assets is taken by intersection points of vehicle paths, and only one vehicle can occupy a path at any one time.

\subsection*{Quantum implementations}
There are many ways to compute graph coloring using neutral-atom quantum computers, by computing the maximum independent set of the mapped coloring graph. The most natural implementation is adiabatic quantum computing \cite{Albash2018}, where the state of a system is slowly evolved to follow the ground state of some time-dependent Hamiltonian. The ground state of the Hamiltonian may be chosen to encode the maximum independent set of a graph in an extremely natural manner using the Rydberg blockade and detuning fields, and so measuring the final wavefunction of the time-evolved system encodes the maximum independent set of the problem. If the problem cannot be encoded into a unit disk graph as may be the case for non-geometric graphs, it can alternatively be reduced to a unit disk graph using extra ancillary qubits~ \cite{nguyen2023}. Beyond adiabatic protocols, variational protocols such as the QAOA \cite{farhi2014quantum}, general variational ansatz \cite{Cerezo2021}, and quantum machine learning (QML) \cite{Biamonte2017} may be used to generate wavefunctions which encode the maximum independent set with an efficacy that may outperform classical computers. Indeed, a demonstration at Harvard in collaboration with QuEra shows that neutral-atom quantum processors can solve the MIS problem with a superlinear quantum speedup compared to a class of generic classical algorithms~\cite{ebadi2022quantum}.

\quad

\newpage

%
%
%
%
%
%

\begin{table}[h]\label{beyond_independent_set}
\begin{tabular}{|c|}
\hline
\parbox{\linewidth}{\vspace{-5mm}
\section{Solving problems beyond independent set using reductions and mappings}\label{sec:beyondUDG}
\vspace{-3mm}}\\ \hline
\quad\includegraphics[width=0.95\linewidth]{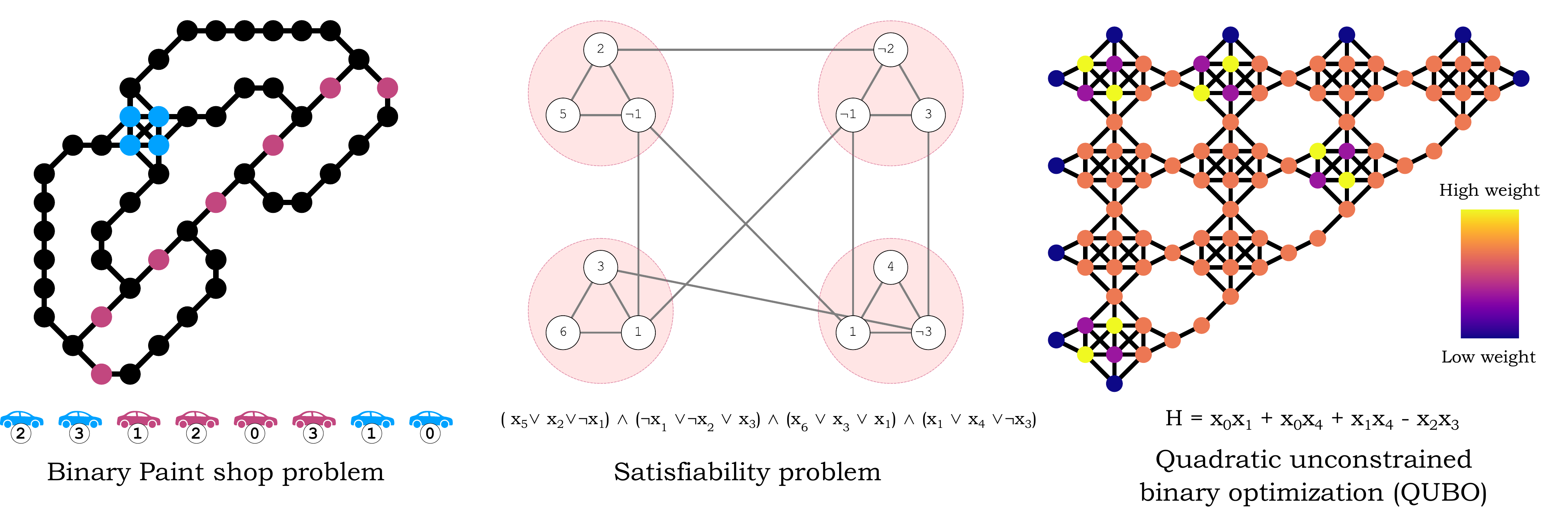}\\

\parbox{0.95\linewidth}{\justify{Other hard combinatorial optimization problems can be solved using solutions from maximum independent set problems using reductions and mappings. Some example mappings are shown above.\\
\textbf{Left} shows an example problem mapping from the \emph{binary paint shop problem}. In this problem \cite{Streif_2021}, pairs of cars enter a ``paint shop" in random order and need to be painted. Each pair must be painted opposite colors, and the shop must decide which color to paint the first of each pair. There is a penalty to switching the color, and so the optimal choice of coloring minimizes the number of switches. This optimal coloring may be encoded in the ground state of an MIS problem on a unit disk graph, as illustrated above.\\
\textbf{Middle} shows an example problem mapping from the \emph{SAT problem}. The decision version is to decide if a logical equation is satisfiable, i.e.~if there is some choice of literals $x_i\in  \{ \texttt{True}, \texttt{False} \}$ such that the equation evaluates to \texttt{True}. In conjunctive normal form, each clause (of the form A, B, or C) is represented as a clique of an independent set graph, while each node represents a literal or its inverse. Edges connect inverse nodes, and the equation is satisfiable if the independent set size is equal to the number of clauses.\\
\textbf{Right} shows an example problem mapping from the \emph{QUBO problem}. In this problem, one must find a bitstring $x_i \in \pm 1$ which minimizes or maximizes the total weight of some quadratic polynomial $H$. By choosing weights of a unit disk graph, the optimal bitstring $x_i$ is encoded in the maximum weighted independent set
(\textit{manuscript in preparation}). }}
\\\quad\\\hline
\end{tabular}
\vspace{-5mm}
\end{table}
A powerful fact about the maximum independent set problem is that it is \texttt{NP-complete}, even on unit disk graphs \cite{Pichler2018MIS, pichler2018Complexity}. This property means that solvers for MIS can be used to tackle any other problem in \texttt{NP} by \emph{reducing} that problem to MIS. This is usually done through some polynomial reduction, where the MIS of some large graph encodes the solution to, say, the satisfiability problem. This fact lets us potentially solve a much broader class of general \texttt{NP} problems on a quantum device, by mapping one problem to a particular choice of graph.

One natural example of this mapping has already been shown in section \ref{sec:graph-coloring}, which reduces graph coloring to a maximum independent set problem. In that example, extra vertices are added to each vertex of the original graph, where each new vertex represents a color. If a vertex in each cluster is part of the independent set, the graph is $k$-colorable, which is the decision version of the \texttt{NP} problem. Similarly, SAT, a paradigmatic \texttt{NP} problem, can be mapped by clustering clauses, as shown above.

The challenge is then in mapping problems from their natural form to a unit disk graph that can fit onto today's neutral-atom quantum computers. While standard reductions exist between problems \cite{Karp1972}, as well as reductions of arbitrary problems directly to unit disk independent set \cite{nguyen2023}, it is fruitful to find reductions from the original problem directly to a problem-specific MIS encoding on unit disk graphs. In this way, by being aware of the substructure of the problem, a more efficient and robust encoding might be constructed that takes advantage of today's neutral-atom quantum computers. For example, the binary paint shop problem (left) may be mapped to a QUBO problem (right) \cite{Streif_2021} using the arbitrary reduction, or more efficiently directly into a unit disk mapping (left) at reduced overhead in atom number and layout area.

In summary, neutral-atom quantum computers are not only well-suited to solve the class of independent set problems directly, but a number of other useful combinatorial optimization problems can be mapped into independent set problems with very little overhead, taking advantage of the specific structure of the problems, so today's neutral-atom quantum computers can be used to tackle those problems too. 

\newpage

%
%
%
%
%

	\bibliography{citationlist}

\end{document}